\documentclass[sigconf,nonacm]{acmart}
\setcopyright{none}
\renewcommand\footnotetextcopyrightpermission[1]{}

\usepackage{booktabs,tabularx,longtable,array}
\usepackage{enumitem}
\usepackage{multirow}
\usepackage{xspace}
\usepackage{graphicx}
\usepackage{multibib}
\newcites{app}{Appendix References}
\graphicspath{{figures/}}

\newcommand{\same}{\textcolor{blue}}
\newcommand{\zero}{\textcolor{gray}}
\newcommand{\opp}{\textcolor{red}}

\title{Bias Cascade: Understanding the Gap Between Knowledge Alignment and Effect Replication in LLM-based Social Simulation}
\author{Pujun Zheng}
\affiliation{%
  \institution{School of Economics and Management, East China Normal University}
  \city{Shanghai}
  \country{China}}
\email{pjzheng@stu.ecnu.edu.cn}

\author{Chengmu Wang}
\affiliation{%
  \institution{School of Economics and Management, East China Normal University}
  \city{Shanghai}
  \country{China}}
\email{chengmu.wang@stu.ecnu.edu.cn}

\author{Wanying Ren}
\affiliation{%
  \institution{School of Statistics, East China Normal University}
  \city{Shanghai}
  \country{China}}
\email{wyren@stu.ecnu.edu.cn}

\author{Jiacheng Yao}
\affiliation{%
  \institution{School of Economics and Management, East China Normal University}
  \city{Shanghai}
  \country{China}}
\email{jcyao@stu.ecnu.edu.cn}

\author{Peizhou Yu}
\affiliation{%
  \institution{School of Economics and Management, East China Normal University}
  \city{Shanghai}
  \country{China}}
\email{pzyu@stu.ecnu.edu.cn}

\author{Xiaoqian Zhang}
\affiliation{%
  \institution{School of Economics and Management, East China Normal University}
  \city{Shanghai}
  \country{China}}
\email{xqzhang@stu.ecnu.edu.cn}

\author{Xiongyao Zha}
\affiliation{%
  \institution{School of Economics and Management, East China Normal University}
  \city{Shanghai}
  \country{China}}
\email{xyzha@stu.ecnu.edu.cn}

\author{Guoxiu He}
\correspondingauthor
\thanks{\ding{41} Corresponding author.}
\affiliation{%
  \institution{School of Economics and Management, East China Normal University}
  \city{Shanghai}
  \country{China}}
\email{gxhe@fem.ecnu.edu.cn}
\def\shortauthors{Zheng et al.}
\acmYear{2027}
\copyrightyear{2027}
\date{}

\begin{document}

\begin{abstract}
Large language models (LLMs) are increasingly used to simulate participants in social experiments. Existing research has primarily focused on aligning social knowledge and isolated responses, while overlooking whether simulations can reproduce the collective effects of social interventions. This distinction is critical due to a failure mechanism we call a \emph{bias cascade}: similar individual states can lead to different actions, while agreement on isolated actions can still produce divergent outcomes through dynamic interactions. As a result, models may produce accurate social judgments yet fail to reproduce the outcomes of classic human studies. To investigate this problem empirically, we introduce \emph{REFLEX} (Reproducing Effects of Foundational Literature EXperiments), a benchmark of 30 counterfactual social experiments grounded in classic human studies that compares knowledge judgment with effect simulation. Knowledge judgment directly predicts the direction of an intervention's effect, whereas effect simulation fixes the scenario and participants while varying one prespecified social factor, with agents generating actions from local observations that are aggregated into macro-level average treatment effects. Compared with human-study findings, the best-performing model achieves net directional scores of 0.847 for knowledge judgment and 0.453 for effect simulation. Moreover, longer causal chains are associated with poorer effect replication, consistent with the bias-cascade hypothesis. These findings reveal a substantial gap between knowledge alignment and effect replication, showing that current LLMs remain inadequate for reliably simulating collective human behavior.
\end{abstract}

\ccsdesc[500]{Computing methodologies~Modeling and simulation}

\keywords{large language models; social simulation; knowledge alignment; treatment-effect direction}

\maketitle

\section{Introduction}

Large language models (LLMs) are increasingly used to simulate participants in social experiments, supporting institutional comparison, intervention design, and experimental rehearsal \cite{hu2026simbench,liu2026humanstudy,kolluri2025finetuning}. These applications require simulations not only to capture how humans respond to changing social conditions, but also to reproduce the collective effects observed in human experiments.

\begin{figure}
	\centering
	{\includegraphics[width=1\columnwidth]{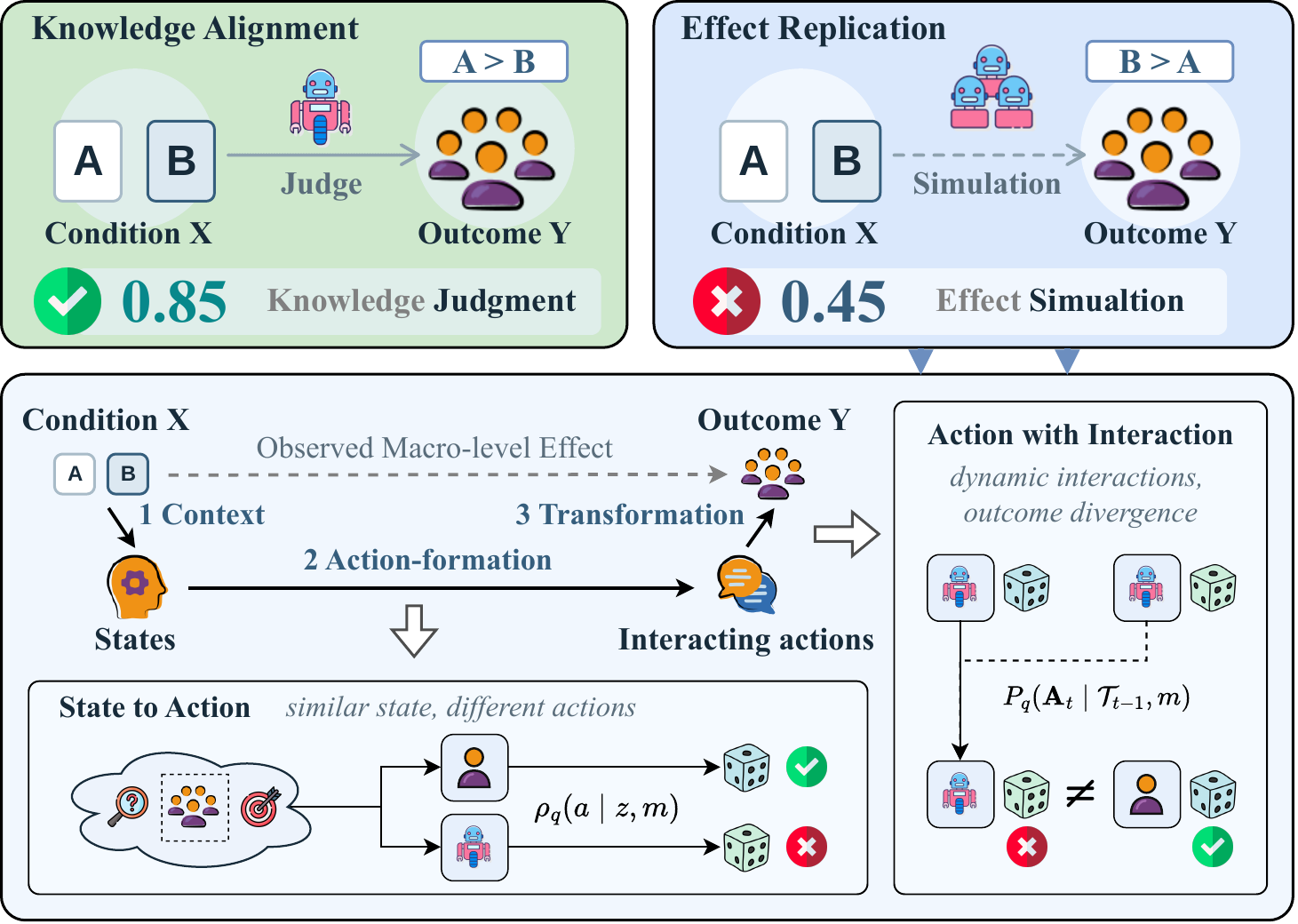}}
	\caption{From knowledge judgments to effect simulation. The top panel compares judgments and simulation. The bottom panel shows the situational, action-formation, and transformation mechanisms that a social simulation must align, with participants interacting in different environments and simulated effects compared with human-study references.}
	\label{fig:teaser}
\end{figure}

Existing approaches broadly follow two strategies to enhance model capabilities: aligning social knowledge through demonstrations, human preferences, simulated social interactions, and structured knowledge tasks \cite{ouyang2022training,liu2024training,choi2023llms,li2026also}; and simulating participants based on roles, population profiles, survey responses, or personal self-reports to approximate the judgments of target individuals or populations \cite{argyle2023out,park2024llm,kolluri2025finetuning}.

However, these approaches leave important gaps. First, from a theoretical perspective, existing methods emphasize isolated decisions or individual attributes while underrepresenting dynamic interaction mechanisms \cite{jang2026instruction,li2026ai}. Faithful simulation requires more than plausible individual judgments. In the social sciences, empirical associations between macro-level variables are not sufficient explanations, because macro-level outcomes can emerge from micro-level processes \cite{schelling2006micromotives,granovetter1978threshold}. To analyze effects faithfully, agents must change their actions in a dynamic environment, rather than make one-way, static choices as in many studies. Second, from a evaluation perspective, model evaluation often prioritizes agreement with observed responses over fidelity to causal effects \cite{hu2026simbench,bisbee2024synthetic,li2026statistical}. Testing whether a social factor changes a collective outcome requires interventions that vary the focal factor while holding other conditions fixed \cite{holland1986statistics,goyal2026plausible}. The resulting treatment effect, including its direction, provides a more direct test of effect replication than agreement on isolated responses. Optimizing only for response-level agreement risks rewarding correlational shortcuts that fail under intervention.

We use Coleman's boat from the social sciences to provide a theoretical account: macro-level conditions shape individual states through a situational mechanism, these states generate actions through an action-formation mechanism, and interacting actions produce macro-level outcomes through a transformation mechanism \cite{coleman1990foundations,hedstrom1998social}. This framework reveals two potential sources of mismatch. At the action-formation stage, a model may reproduce human states but apply a different decision rule, resulting in different actions. At the transformation stage, even agreement on isolated actions may still lead to different feedback when agents interact with one another. Errors across these stages can propagate and accumulate, causing simulated effects to diverge from human findings. We call this failure a \emph{bias cascade}.

To test this hypothesis empirically, we introduce \emph{REFLEX} (Reproducing Effects of Foundational Literature EXperiments), a counterfactual social experiment benchmark that tests whether model simulations reproduce intervention effects documented in classic social science studies. REFLEX compares model performance in two settings. In \textit{knowledge judgment}, models directly predict the direction of an intervention's effect in a given social scenario. In \textit{effect simulation}, each of the 30 tasks constructs intervention conditions from a shared scenario, varying one prespecified social factor while holding participant configurations and all other experimental settings identical. Simulated participants act on locally available observations, and their actions are aggregated into macro-level outcomes. We estimate the treatment effect as the mean outcome difference between conditions and compare its direction with the corresponding human-study finding. 

The best-performing model achieves net directional scores of 0.847 for knowledge judgment and 0.453 for effect simulation. The lower simulation score reveals substantial limitations in reproducing human experimental effects, while the gap between evaluation modes shows that accurate social judgments do not reliably translate into faithful simulations. Longer causal chains are also associated with poorer effect replication, consistent with the proposed bias-cascade account.

Our contributions are as follows:

\begin{enumerate}[leftmargin=*]

	\item \textbf{Theoretical analysis.} We formalize the social simulation process, showing that neither state matching nor isolated action matching guarantees faithful macro-level effects.
	
	\item \textbf{Benchmark construction.} We introduce \emph{REFLEX}, a benchmark of 30 counterfactual social experiments for evaluating whether simulated treatment effects match the directions documented in human studies.
	
	\item \textbf{Experimental validation.} We quantified the gap between knowledge alignment and effect replication, and found consistent and significant differences among different LLMs.
	
	\item \textbf{Empirical diagnosis.} We identify an association between longer causal chains and poorer effect replication, providing evidence consistent with bias cascades.

\end{enumerate}  

\section{Related Work}
\subsection{LLM-Based Social Simulation}
Research on LLM-based social simulation has progressed along two main directions: simulation frameworks and agent construction. Frameworks model ongoing social interactions by defining agents' backgrounds, memories, and goals, and incorporating environmental feedback. Representative systems include Generative Agents \cite{park2023generative}, which supports memory and planning; Concordia \cite{vezhnevets2023generative}, which separates agent decision-making from environmental execution; AgentSociety \cite{piao2025agentsociety}, which supports large-scale social experiments; and OdysSim \cite{zhou2026odyssim}, which builds foundation models for human behavior simulation. Agent construction methods use demographic attributes, interview data, or responses from human experiments to create silicon samples or individual agents \cite{argyle2023out,park2024llm,binz2025foundation,kolluri2025finetuning,du2026twinvoice}. Together, these approaches provide platforms for social simulation and ways to ground agents in individual-level attributes. Rather than scaling platforms or matching static responses, we examine whether interacting agents can faithfully reproduce collective treatment effects under controlled interventions.

\subsection{Validity Challenges in Social Simulation}
Prior work highlights that social simulation validity cannot be assessed solely from final outcomes or average responses. Errors may arise from the simulation process, the response distribution, or the experimental setup. At the process level, a model may reach the correct outcome through an incorrect generative path or mechanism \cite{lee2026slalom,pandey2026reason,zhao2026mechanism}. Scheduling, information exposure, and environmental update rules may also shape collective outcomes more strongly than communication among agents \cite{li2026ai}. At the distributional level, synthetic responses may approximate the human mean while compressing variance, distorting relationships between variables, or underrepresenting minority opinions \cite{bisbee2024synthetic,bae2026marginal,qin2026restoring,li2025chatgpt}. Models may also describe a target distribution yet fail to generate behavior consistent with it across repeated calls \cite{jang2026instruction}. At the experimental level, prompts, role assignments, and interaction protocols can substantially affect group outcomes \cite{ye2026stop,michailidis2026caricature,aizaz2026analysing}. Results may also depend on simulation configuration, model scale, and action-selection methods \cite{cummins2026threat,ziems2026scaling,buitragolopez2026llm}. More broadly, counterfactual estimates depend on the simulator, and their real-world relevance hinges on its fidelity \cite{goyal2026plausible}. Although these studies identify diverse validity risks, they do not explain why models with human-aligned social knowledge still fail to reproduce experimental effects at the macro level. We address this gap by analyzing how mismatches in individual decision-making and multi-agent interaction propagate and accumulate across stages.

\subsection{Benchmarks for Social Simulation}
Existing benchmarks assess group response distributions, individual test-retest consistency, multi-turn social interaction, and simulation processes and trajectories \cite{santurkar2023opinions,hu2026simbench,toubia2025database,zhou2024sotopia,lee2026slalom,wang2025sociobench}. They cover response similarity, interaction capabilities, and process plausibility, but do not directly test whether intervention effects can be reproduced through actual actions. Related work has extended evaluation beyond response similarity to effect reproduction. Studies have tested whether LLMs can reproduce classic human experiments \cite{aher2023using} and estimated treatment effects from isolated simulated responses \cite{ashokkumar2026large}. Other work incorporates experimental materials, conditions, and statistical procedures into end-to-end evaluations \cite{liu2026humanstudy}, or separately measures behavioral distribution fit and contextual adaptation \cite{tareaf2026benchmarking}. Selecting models, prompts, or target populations based on individual response errors has also not consistently reduced treatment effect errors \cite{li2026statistical}. Nevertheless, most evaluations rely on static responses or treat distributional fit as a proxy for causal fidelity. Few use counterfactual simulations to test whether social knowledge translates into emergent group outcomes. Our benchmark directly compares models' directional judgments about interventions with their ability to reproduce those treatment effects through interactive simulation.

\section{Theoretical Analysis and Benchmark Design}
\subsection{Bias Cascade Revealed by Coleman's Boat}

Our study addresses the following question: Why might existing LLMs, despite possessing relevant knowledge and understanding the consequences of human behavior under certain conditions, still fail to reproduce outcomes observed in real human societies? We use a \emph{bias cascade} to explain this failure. The upper panel of Figure~\ref{fig:teaser} illustrates this problem.

We use Coleman's boat to analyze its underlying causes \cite{coleman1990foundations,hedstrom1998social}, drawing on a classic framework in social theory. As shown in the bottom panel of Figure~\ref{fig:teaser}, the relationship between a macro-level condition $X$ and a macro-level outcome $Y$ can be decomposed into a mechanism chain that moves from the macro level to the micro level and back. The situational mechanism explains how institutions, resources, and information structures shape individuals' circumstances. The action-formation mechanism explains how individuals choose actions in response to those circumstances. The transformation mechanism explains how these actions, through interdependence and environmental feedback, produce macro-level outcomes.

With sufficient pretraining and post-training alignment, existing LLMs are already close to humans in many value judgments, in their understanding of social norms, and in their tendencies in routine behavior \cite{dillion2023ai}. We therefore introduce an assumption about how models understand social situations as the starting point for our formal analysis:

\begin{quote}
\textit{\textbf{Assumption.} LLMs understand social situations. That is, given a specified type of social situation, their ability to understand agents' micro-level states, namely beliefs and goals, is comparable to that of humans.}
\end{quote}

Let $q\in\{H,L\}$ index humans and models, respectively, and let $m\in\{0,1\}$ denote the externally assigned macro-level control and treatment conditions. Throughout, conditioning on $m$ refers to the distribution in the corresponding intervention world. Let $\mathcal T$ denote a complete interaction trajectory, and let $g$ be an outcome function shared by both types of systems. The target treatment effect is
\begin{equation}
\Delta_q=\mathbb E_q[g(\mathcal T)\mid m=1]-\mathbb E_q[g(\mathcal T)\mid m=0].
\end{equation}

The \textbf{first source of bias} lies between beliefs and actions. Let $z$ collectively represent a given individual's circumstances, beliefs, and goals. Let $Q_q(z\mid m)$ denote the distribution of the individual's micro-level state, and let $\rho_q(a\mid z,m)$ denote the action-selection rule given that state. These two components determine the distribution of individual actions. For simplicity, we consider discrete states:
\begin{equation}
P_q(A=a\mid m)=\sum_z\rho_q(a\mid z,m)\cdot Q_q(z\mid m).
\end{equation}

This equation distinguishes an individual's state from how they act on that basis. Even under an assumption stronger than state matching, namely that the joint distribution of states $Q_q$ is matched, actions still depend on the mapping from states to choices, $\rho_q$. The same circumstances, beliefs, and goals do not uniquely determine a decision rule or its execution. In realistic settings, models and humans may differ in this mapping: a model may describe a target distribution without reliably sampling from it across repeated actions \cite{jang2026instruction}. Model-generated actions may therefore diverge from actual human behavior.

\begin{proposition}[State matching does not guarantee action matching]
Identical micro-level states under both intervention conditions are not sufficient to guarantee identical actions:
\begin{equation}
\begin{gathered}
Q_L(\cdot\mid m)=Q_H(\cdot\mid m),\quad\forall m\\
\not\Rightarrow\quad P_L(A\mid m)=P_H(A\mid m),\quad\forall m.
\end{gathered}
\end{equation}
\end{proposition}

\paragraph{Constructive proof} Fix an individual's other circumstances and goals. Suppose both humans and models believe that cooperation is beneficial with probability $1/4$ when $m=0$ and $3/4$ when $m=1$. Humans cooperate only when they believe cooperation is beneficial, whereas the model always cooperates, reflecting a prosocial tendency induced by preference alignment. The belief distributions are identical for the two systems under each condition, but the human cooperation rate rises from $1/4$ to $3/4$, while the model's rate remains at $1$. If the outcome is the group-average cooperation rate, then $\Delta_H=1/2$ and $\Delta_L=0$, so the effects differ.

The \textbf{second source of bias} lies between marginal action distributions and interaction processes. Let $A_{i,t}$ denote the action of individual $i$ at time $t$, $\mathbf A_t=(A_{1,t},\ldots,A_{n,t})$ denote the actions of all individuals, and $S_t$ denote the environmental state. The individual marginal distribution $P_q(A_{i,t}\mid m)$ gives the probability that a given individual chooses a particular action at time $t$. In a dynamic environment, the relevant object is the complete interaction trajectory $\mathcal T=(S_0,\mathbf A_1,S_1,\ldots,\mathbf A_T,S_T)$. Let $\mathcal T_{t-1}$ denote the full history before the actions in round $t$, that is, the trajectory prefix up to $S_{t-1}$, with $\mathcal T_0=(S_0)$. By the chain rule of probability,
\begin{equation}
\begin{aligned}
P_q(\mathcal T\mid m)
={}&P_q(S_0\mid m)\\
&\prod_{t=1}^{T}\Big[
P_q(\mathbf A_t\mid\mathcal T_{t-1},m)\cdot P_q(S_t\mid\mathcal T_{t-1},\mathbf A_t,m)\Big].
\end{aligned}
\end{equation}

This equation shows that even if an LLM closely matches each individual's action probabilities, the dependencies among individuals may still differ. The first conditional term in the product describes how individuals jointly act in response to history, while the second describes how their actions change the environment. Crucially, the trajectory distribution also captures dependencies among individuals, responses to prior history, and feedback between actions and the environment. Individual action marginals do not determine these interaction structures. Classic analyses show that individual-level regularities do not automatically determine collective outcomes \cite{schelling2006micromotives,granovetter1978threshold}. Consequently, even small discrepancies can produce substantial outcome divergence in systems shaped by complex interactions.

\begin{proposition}[Matching action marginals does not guarantee matching interaction effects]
Even if individuals' action distributions match exactly at every time point and under every intervention condition, the distributions over complete trajectories may still differ:
\begin{equation}
\begin{gathered}
P_L(A_{i,t}\mid m)=P_H(A_{i,t}\mid m),\quad\forall i,t,m\\
\not\Rightarrow\quad P_L(\mathcal T\mid m)=P_H(\mathcal T\mid m),\quad\forall m.
\end{gathered}
\end{equation}
\end{proposition}

Therefore, even with the same outcome function $Y=g(\mathcal T)$, matching the marginals alone does not guarantee that the macro-level outcome distributions $P_q(Y\mid m)$ agree. For measures that depend on action combinations or interaction histories, the average treatment effects may even have opposite signs. We provide a construction involving actual information exchange below.

\paragraph{Constructive proof} Consider a two-person, two-round coordination task in which actions are either $0$ or $1$. In both systems, the two individuals independently choose their first-round actions with equal probability. Individual 1 always repeats their first-round action in the second round. The only difference between conditions is whether individual 2 observes individual 1's first-round action before the second round. Under the control condition $m=0$, the action is not disclosed, and individual 2 repeats their own first-round action. Under the treatment condition $m=1$, the action is disclosed: the human individual 2 follows the observed action, whereas the model chooses the opposite action. The systems differ only in how participants respond to the observation. Under every condition and at every time point, each individual chooses $1$ with probability $1/2$, so all marginal distributions match. However, if the outcome is whether coordination succeeds in the second round, then $\mathbb E_H[Y\mid m=0]=1/2$, $\mathbb E_H[Y\mid m=1]=1$, and $\Delta_H=+1/2$, whereas $\mathbb E_L[Y\mid m=0]=1/2$, $\mathbb E_L[Y\mid m=1]=0$, and $\Delta_L=-1/2$. The individual action probabilities match across the two systems, but information disclosure promotes coordination in one and anti-coordination in the other, reversing the macro-level effect.

Together, these results illustrate the difficulty of simulation. Existing models may be well aligned with relevant social knowledge yet still struggle to reproduce the effects observed in real human societies. The central challenge lies in these two sources of mismatch in the simulation process. Discrepancies between models and humans can cascade and amplify across stages, ultimately producing substantial divergence between simulated and actual human outcomes.

\subsection{Counterfactual Task Construction}

We define a counterfactual task as
\begin{equation}
\mathcal P_s=(S_s,m_s,\pi_s,g_s,\mathcal H_s),
\end{equation}
where $S_s$ is the base scenario and initial-condition distribution for task $s$, $m_s\in\mathcal M_s$ denotes the active manipulation across experimental conditions with $\mathcal M_s=\{0,1\}$ representing control and treatment in standard two-condition tasks, $\pi_s$ is the execution protocol held fixed apart from that manipulation, $g_s$ is a prespecified behavioral readout function, and $\mathcal H_s$ contains the task's specification details, including the treatment definition, outcome variable, effect contrast, reference direction, and scope of applicability.

The minimal-manipulation constraint requires experimental conditions to differ only in the prespecified exogenous treatment factor. It does not require all post-treatment variables to remain unchanged. A change in penalty can alter the action space and expected payoffs; a change in information allocation can alter what participants say; and a change in network structure can alter exposure pathways. These changes may be mechanisms through which the treatment produces its effect, so fixing them after treatment would block the effect being measured. Each task must therefore distinguish explicitly among fixed initial factors, the active manipulation itself, and subsequent state changes caused by the manipulation.

All conditions within a task use the same model settings, agent specifications, and execution protocol. Environment seeds are paired, and sessions use independent contexts and private memories. Paired environment seeds control only implementation-level initial randomness; model sampling noise and other within-session randomness remain independent. Treatment effects are estimated as differences in condition means across repeated experimental units, not inferred from a single output.

Let $u$ denote an experimental unit, and let $\mathcal T_{s,u}(m)$ be the complete trajectory produced by unit $u$ in task $s$ under condition $m\in\mathcal M_s$. Define the outcome variable as
\begin{equation}
Y_{s,u}(m)=g_s\!\left(\mathcal T_{s,u}(m)\right).
\end{equation}
For multi-agent tasks, the full group, network, or interaction sequence constitutes one experimental unit. Interactions among group members are included in the unit-level outcome, and members must not be treated as independent treatment samples.

For task $s$ and seed $r$, let $\overline Y_{s,r}(m)$ be the mean of the primary outcome across repeated experimental units under condition $m$. For a standard two-condition task, we use
\begin{equation}
\widehat{\Delta}_{s,r}
=\overline Y_{s,r}(1)-\overline Y_{s,r}(0).
\end{equation}
For tasks with multiple intervention conditions, the treatment effect is calculated as a prespecified contrast:
\begin{equation}
\widehat{\Delta}_{s,r}
=\sum_{m\in\mathcal M_s} w_{s,m}\overline Y_{s,r}(m),
\qquad
\sum_{m\in\mathcal M_s} w_{s,m}=0,
\end{equation}
where the weights $w_{s,m}$ are fixed before the experiment and sum to zero.

Let $h_s\in\{-1,+1\}$ be the reference direction for task $s$, and let $\tau=10^{-10}$ be a small threshold. Define the direction-adjusted effect as
\begin{equation}
e_{s,r}=h_s\widehat{\Delta}_{s,r}.
\end{equation}
A result is classified as aligned if $e_{s,r}>\tau$, reversed if $e_{s,r}<-\tau$, and null if $|e_{s,r}|\leq\tau$.

\subsection{Evaluation Metrics}

Let $S$ be the number of tasks and $R$ be the number of valid seeds per task. We report five evaluation metrics.

\textbf{Same Direction (same)} is the binary variable
\begin{equation}
\mathrm{same}_{s,r} = \mathbf{1}\{e_{s,r}>\tau\}.
\end{equation}

\textbf{Standardized Treatment Effect (std\_effect)} is the continuous variable
\begin{equation}
\operatorname{std\_effect}_{s,r} = \frac{e_{s,r}}{\operatorname{sd}_{s}(e)},
\end{equation}
where $\operatorname{sd}_{s}(e)$ denotes the standard deviation of $e_{s,r}$ within task $s$.

\textbf{Directional Alignment Rate (DAR)} measures the proportion of valid results that are aligned:
\begin{equation}
\mathrm{DAR}
=
\frac1S\sum_{s=1}^{S}
\frac{N_{s,\mathrm{same}}}{R},
\end{equation}
where $N_{s,\mathrm{same}}$ is the number of valid results for task $s$ satisfying $e_{s,r}>\tau$. DAR lies in $[0,1]$.

\textbf{Directional Reversal Rate (DRR)} measures the proportion of valid results that are reversed:
\begin{equation}
\mathrm{DRR}
=
\frac1S\sum_{s=1}^{S}
\frac{N_{s,\mathrm{opposite}}}{R},
\end{equation}
where $N_{s,\mathrm{opposite}}$ is the number of valid results for task $s$ satisfying $e_{s,r}<-\tau$. DRR lies in $[0,1]$.

\textbf{Net Directional Score (NDS)} is defined as the difference between the directional alignment rate and the directional reversal rate:
\begin{equation}
\mathrm{NDS}=\mathrm{DAR}-\mathrm{DRR}.
\end{equation}
NDS lies in $[-1,1]$ and measures the net directional response: a score of $1$ indicates that all valid results are aligned, whereas $-1$ indicates that all valid results are reversed.

\subsection{REFLEX Benchmark Design}

\paragraph{Task design} \emph{REFLEX} comprises 30 counterfactual social simulation tasks, grouped into six mutually exclusive social science domains: behavioral and experimental economics; political economy and public governance; social psychology and group processes; networks, communication, and collective intelligence; markets, reputation, and labor; and culture, norms, and evolution. Interdisciplinary studies are assigned to a single category based on their primary research question.

In \textit{effect simulation}, paired runs of the same experimental unit across conditions share agent profiles and environment instances, differing only in the prespecified social factor. Agents persist across multiple rounds, can access only locally visible information, and do not see paper titles, expected effect directions, other agents' private information, or global states. Agents can affect the environment through structured actions, so their actions coevolve with the overall environment. Scoring is strictly based on actual actions, such as contributions, penalties, migration, link formation, offers, and votes, rather than general descriptions of human behavior or textualized trajectories. For comparison, \textit{knowledge judgment} directly queries models to predict the treatment effect direction from structured descriptions of the study's scenario and intervention, without multi-agent execution.

\paragraph{Participant profiles} The platform uses structured heterogeneous profiles to initialize simulated participants. In survey-calibrated mode, 2,185 complete adult records are selected from the SAPA project data. Records are sampled at the individual level, preserving the joint structure of Big Five personality traits, age range, and education. The sample has a mean age of 30.3 years and an average education level equivalent to college.

\section{Experiment}\label{sec:results}

\begin{table*}[t]
	\centering\small
	\caption{Knowledge judgment and effect simulation across the eight models. Entries in the composition columns are + / 0 / - (aligned / null / reversed). K-E is the difference between knowledge judgment NDS and effect simulation NDS. Brackets give bootstrap 95\% confidence intervals.}
	\label{tab:main}
	\begin{tabular*}{\textwidth}{@{\extracolsep{\fill}}lccccccccc@{}}
		\toprule
		Model & \multicolumn{4}{c}{Knowledge Judgment} & \multicolumn{4}{c}{Effect Simulation} & K-E [95\% CI]\\
		\cmidrule(lr){2-5}\cmidrule(lr){6-9}
		& + / 0 / - & DAR $\uparrow$ & DRR $\downarrow$ & NDS $\uparrow$
		& + / 0 / - & DAR $\uparrow$ & DRR $\downarrow$ & NDS $\uparrow$ & \\
		\midrule
		GLM-5.3 & \same{27.6}/\zero{0.2}/\opp{2.2} & \textbf{0.920} & \underline{0.073} & \textbf{0.847} & \same{19.8}/\zero{4.0}/\opp{6.2} & \textbf{0.660} & \textbf{0.207} & \textbf{0.453} & 0.393 [0.113, 0.593]\\
		Claude-Sonnet-5 & \same{26.2}/\zero{1.6}/\opp{2.2} & \underline{0.873} & \textbf{0.073} & \underline{0.800} & \same{17.6}/\zero{3.6}/\opp{8.8} & 0.587 & 0.293 & 0.293 & 0.507 [0.219, 0.747]\\
		Qwen3.8-27B & \same{25.6}/\zero{1.4}/\opp{3.0} & 0.853 & 0.100 & 0.753 & \same{17.4}/\zero{3.8}/\opp{8.8} & 0.580 & 0.293 & 0.287 & 0.466 [0.213, 0.720]\\
		Kimi-K2.6 & \same{25.4}/\zero{0.6}/\opp{4.0} & 0.847 & 0.133 & 0.713 & \same{16.4}/\zero{3.2}/\opp{10.4} & 0.547 & 0.347 & 0.200 & 0.513 [0.180, 0.747]\\
		DeepSeek-V4.1-Flash & \same{25.0}/\zero{1.2}/\opp{3.8} & 0.833 & 0.127 & 0.707 & \same{16.6}/\zero{4.6}/\opp{8.8} & 0.553 & 0.293 & 0.260 & 0.447 [0.173, 0.667]\\
		MiniMax-M2.7 & \same{25.0}/\zero{0.4}/\opp{4.6} & 0.833 & 0.153 & 0.680 & \same{18.2}/\zero{2.8}/\opp{9.0} & \underline{0.607} & 0.300 & \underline{0.307} & 0.373 [0.060, 0.587]\\
		GPT-6-Luna & \same{24.4}/\zero{0.6}/\opp{5.0} & 0.813 & 0.167 & 0.647 & \same{15.8}/\zero{6.2}/\opp{8.0} & 0.527 & \underline{0.267} & 0.260 & 0.387 [0.040, 0.647]\\
		Qwen3.6-35B-A3B & \same{22.8}/\zero{0.4}/\opp{6.8} & 0.760 & 0.227 & 0.533 & \same{14.8}/\zero{4.6}/\opp{10.6} & 0.493 & 0.353 & 0.140 & 0.393 [0.067, 0.667]\\
		\bottomrule
	\end{tabular*}
\end{table*}

\begin{figure}[t]
	\centering
	\includegraphics[width=\linewidth]{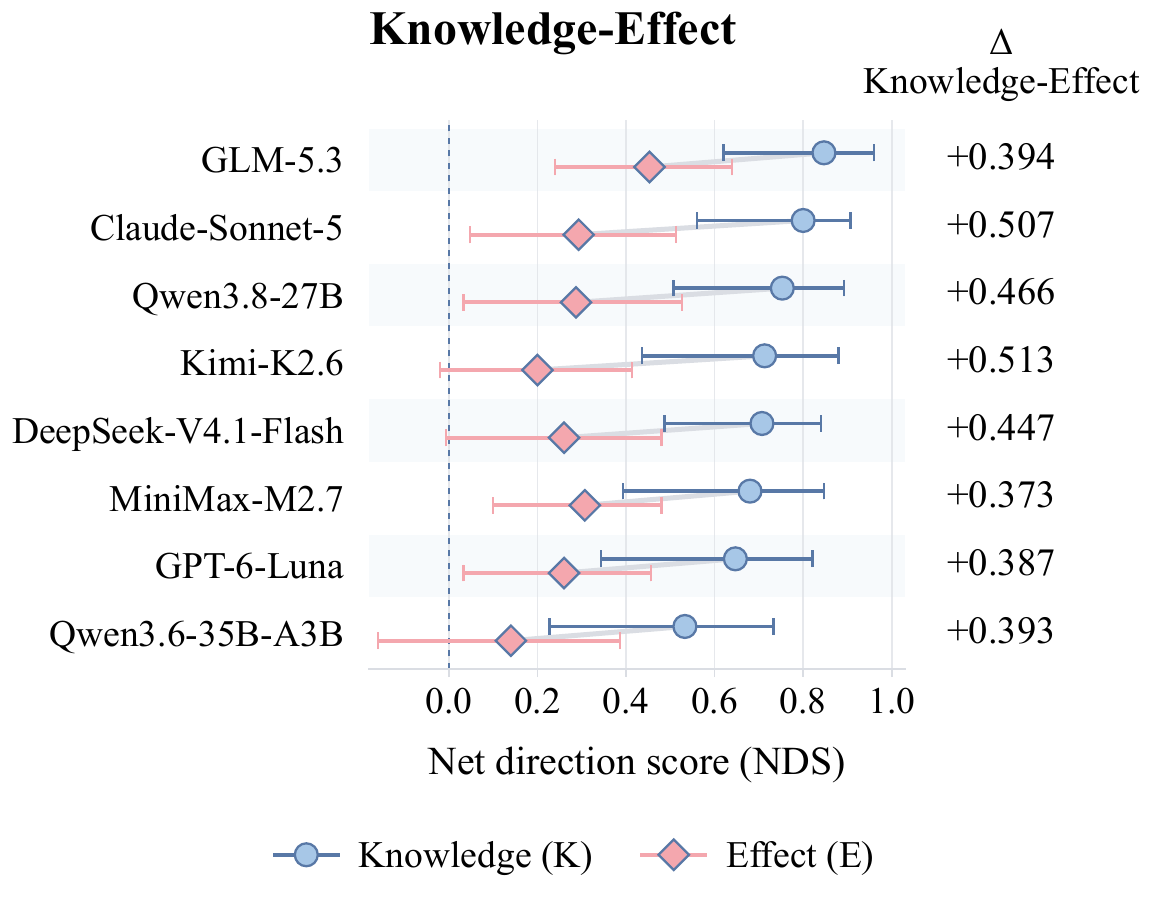}
	\caption{Knowledge judgment and effect simulation NDS for the eight models, ordered from top to bottom by knowledge NDS. NDS takes values in the range$[-1,1]$. The number on the right is their difference. Error bars are 95\% bootstrap confidence intervals. Knowledge NDS exceeds effect NDS for all eight models.}
	\label{fig:nds-dumbbell}
\end{figure}

\subsection{Experimental Setup}
\paragraph{Main experiment} The main experiment evaluates eight open- and closed-source general-purpose commercial models from leading LLM providers: GLM-5.3, MiniMax-M2.7, DeepSeek-V4.1-Flash, Claude-Sonnet-5, Qwen3.8-27B, Kimi-K2.6, GPT-6-Luna, and Qwen3.6-35B-A3B. For each model, we evaluate performance on both effect simulation and knowledge judgment. For effect simulation, we run all 30 \emph{REFLEX} tasks for each model, sampling agent populations with seeds 1 to 5. We use different sampling seeds to vary the sampled population and repeat each experiment five times, while keeping the environment seed fixed at 42. All results are reported with bootstrap confidence intervals.

\paragraph{Ablation experiments} All ablation experiments use GLM-5.3, with the tasks and agent population sampling seeds held consistent with the main experiment. We evaluate three types of ablations. In the information ablation, we provide the model with neutral background information, historical materials from classic studies, or predictions from its own knowledge judgment, and assess whether performance depends on the quality of the input knowledge. In the population-size ablation, we increase the number of agents in each task to 1.5 and 2 times the original setting to test whether model performance depends on population size. In the wording ablation, we change only the shared role-setting instruction and test the model's robustness using several approximate paraphrases.

\subsection{\emph{RQ1}: Is Knowledge Alignment Distinct from Effect Replication?}
Table~\ref{tab:main} presents the full results of the main experiment. The directional alignment rate in effect simulation ranges from 0.493 to 0.660, suggesting that the probability of a model making a directionally correct choice is not negligible. However, errors have substantial consequences for practical policy guidance: an incorrect effect estimate could lead to a misguided policy direction and potentially cause social harm. The directional reversal rate in effect simulation ranges from 0.207 to 0.353. To capture the combined impact of correct and incorrect predictions, we use the net directional score (NDS) as our primary metric. The results show that NDS for effect simulation ranges from 0.140 to 0.453. These scores remain modest and in several cases close to zero, only slightly outperforming random guessing and indicating that models may produce nearly as many incorrect conclusions as correct ones. The low effect simulation score indicates that current methods remain inadequate for simulating human behavior.

We also examine the gap between knowledge judgment and effect simulation, which is illustrated in Figure~\ref{fig:nds-dumbbell}. Consider GLM-5.3, the best-performing model overall in both knowledge judgment and effect simulation. Its NDS is 0.847 for knowledge judgment but only 0.453 for effect simulation. The difference is substantial, at 0.393, with a confidence interval of $[0.113, 0.593]$ that excludes zero, indicating a significant difference between the two settings. We further compare the NDS gap between knowledge judgment and effect simulation across all models. The gap is positive for every model, and each bootstrap confidence interval excludes zero. Thus, the same model exhibits a large difference between the knowledge judgment and effect simulation modes, indicating a fundamental distinction between the two. This result is consistent with our theoretical expectation: key differences in the simulation process prevent this directional knowledge from being carried through into simulated outcomes.

Taken together, these results confirm that knowledge alignment and effect replication differ significantly, and that the knowledge-effect gap is positive for every model.

\subsection{\emph{RQ2}: Is Effect Replication Associated with Model, Information, Population Size, or Wording?}

\paragraph{Information ablation}
We vary the information available to the model at the start of each simulation, using three settings: providing neutral background information about the experiment, providing historical materials related to classic studies, and directly feeding the model's own knowledge judgment answer back into its context. As shown in Table~\ref{tab:information}, NDS is 0.433 with neutral background information and 0.413 with historical materials. Neither score improves on the baseline. Feeding the model its own knowledge judgment answer back into the context raises NDS to 0.520, a modest increase of 0.067 over the baseline. However, the confidence interval, $[-0.027, 0.233]$, still includes zero. The confidence intervals for all three experimental settings include zero, indicating that the differences are not significant. This suggests that even when the relevant content is carefully isolated and background knowledge about the expected effect is explicitly provided to the model, its simulation results do not improve significantly.

Together with the results in Table~\ref{tab:main}, these findings suggest that LLMs are not unaware of what constitutes plausible human behavior, yet simulation fails to reproduce outcomes they themselves explicitly predict. This may seem counterintuitive, as none of the attempted manipulations yields a significant improvement. However, our theoretical analysis points to an explanation: the central reason LLMs fail at social simulation is not insufficient input knowledge or an inability to understand social situations. Rather, models do not capture the actual patterns of human action implicit in that social knowledge, and this error is further amplified through interaction with the social environment.

\begin{table}[ht]
\centering\small
\setlength{\tabcolsep}{1.5pt}
	\caption{Information ablation. The $\Delta$ column reports pointwise 95\% BCa bootstrap confidence intervals ($B=2{,}000$, seed $555$). The baseline NDS is the GLM-5.3 main-experiment effect simulation NDS of 0.453; $\Delta$ is measured against this baseline.}
\label{tab:information}
\begin{tabular*}{\linewidth}{@{\extracolsep{\fill}}lrrr@{}}
\toprule
Configuration & + / 0 / - & NDS $\uparrow$ & $\Delta$ NDS [95\% CI]\\
\midrule
Original & \same{19.8}/\zero{4.0}/\opp{6.2} & 0.453 & ---\\
Neutral & \same{19.6}/\zero{3.8}/\opp{6.6} & 0.433 & $-0.020$ $[-0.180, 0.147]$\\
Historical & \same{19.0}/\zero{4.4}/\opp{6.6} & 0.413 & $-0.040$ $[-0.180, 0.113]$\\
Own judgment & \same{20.8}/\zero{4.0}/\opp{5.2} & 0.520 & $+0.067$ $[-0.027, 0.233]$\\
\bottomrule
\end{tabular*}
\end{table}

\paragraph{Population-size ablation}
We increase the number of agents participating in each task to 1.5 and 2 times the original population size to assess whether model performance remains consistent in larger groups. As shown in Table~\ref{tab:population}, NDS is 0.427 and 0.520 at 1.5 and 2 times the original population size, respectively. These scores represent a decrease of 0.027 and an increase of 0.067 relative to the baseline. The confidence intervals for the differences from the baseline are $[-0.173, 0.133]$ and $[-0.120, 0.267]$, respectively, and both include zero, indicating no significant difference. This suggests that model performance does not depend significantly on the experimental population size.
\begin{table}[ht]
\centering\small
\setlength{\tabcolsep}{1.5pt}
	\caption{Population-size ablation. The $\Delta$ column reports pointwise 95\% BCa bootstrap confidence intervals ($B=2{,}000$, seed $555$). The baseline NDS is the GLM-5.3 main-experiment effect simulation NDS of 0.453; $\Delta$ is measured against this baseline.}
\label{tab:population}
\begin{tabular*}{\linewidth}{@{\extracolsep{\fill}}lrrr@{}}
\toprule
Configuration & + / 0 / - & NDS $\uparrow$ & $\Delta$ NDS [95\% CI]\\
\midrule
Original & \same{19.8}/\zero{4.0}/\opp{6.2} & 0.453 & ---\\
Population $\times1.5$ & \same{19.6}/\zero{3.6}/\opp{6.8} & 0.427 & $-0.027$ $[-0.173, 0.133]$\\
Population $\times2$ & \same{20.6}/\zero{4.4}/\opp{5.0} & 0.520 & $+0.067$ $[-0.120, 0.267]$\\
\bottomrule
\end{tabular*}
\end{table}

\paragraph{Wording ablation}
The wording ablation changes only the shared role-setting instruction and tests the model's robustness using several approximate paraphrases. We evaluate five paraphrases; Table~\ref{tab:wording} reports the two most representative results. The worst-performing paraphrase reduces NDS to 0.220, with a confidence interval of $[-0.440, -0.047]$ that excludes zero, indicating a significant decline in performance. The best-performing paraphrase increases NDS to 0.593, with a confidence interval of $[0.013, 0.287]$ that excludes zero, indicating a significant improvement. The overall range of NDS variation is 0.373. These results show that the model is highly sensitive to minor wording changes, with substantial variation in performance. This indicates that the model is limited not only in performance but also in robustness, which may substantially constrain its use in real-world settings.
\begin{table}[ht]
\centering\small
\setlength{\tabcolsep}{1.5pt}
	\caption{Wording ablation. The $\Delta$ column reports pointwise 95\% BCa bootstrap confidence intervals ($B=2{,}000$, seed $555$). The baseline NDS is the GLM-5.3 main-experiment effect simulation NDS of 0.453; $\Delta$ is measured against this baseline.}
\label{tab:wording}
\begin{tabular*}{\linewidth}{@{\extracolsep{\fill}}lrrr@{}}
\toprule
Configuration & + / 0 / - & NDS $\uparrow$ & $\Delta$ NDS [95\% CI]\\
\midrule
Original & \same{19.8}/\zero{4.0}/\opp{6.2} & 0.453 & ---\\
Rewrite \textit{worst} & \same{16.2}/\zero{4.2}/\opp{9.6} & 0.220 & $-0.233$ $[-0.440, -0.047]$\\
Rewrite \textit{best} & \same{22.2}/\zero{3.4}/\opp{4.4} & 0.593 & $+0.140$ $[0.013, 0.287]$\\
\bottomrule
\end{tabular*}
\end{table}

Taken together, effect replication is significantly associated with model choice and prompt wording, but not with population size or background information. We examine which factor dominates the remaining variation below.

\section{Empirical Analysis}

\subsection{\emph{RQ3}: What Core Factors Shape Model Effect Gaps?}
\begin{figure}[t]
    \centering
    \includegraphics[width=\linewidth]{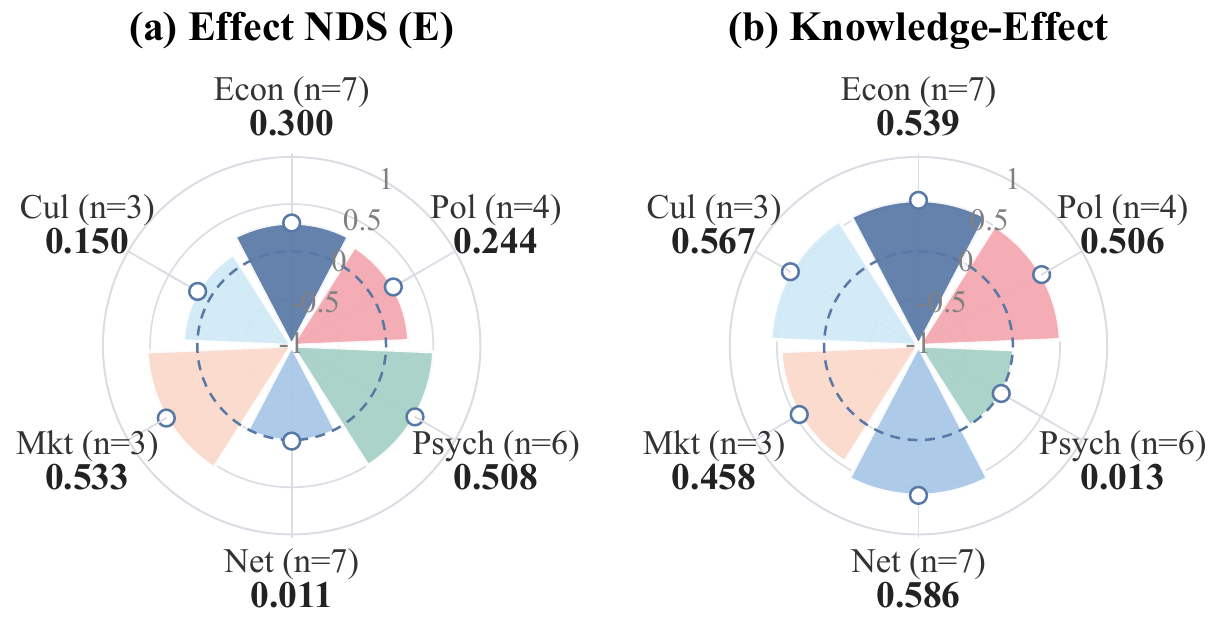}
    \caption{Effect NDS and the knowledge-effect gap (K-E) across the six social science categories. Both panels map the index from $[-1,1]$ to radius $[0,1]$; labels show the original values, the dashed ring denotes zero, the center is $-1$, and the outer ring is $+1$.}
    \label{fig:category-radial}
\end{figure}

\paragraph{Evaluation results by category}
As shown in Table~\ref{tab:category-results} and Figure~\ref{fig:category-radial}, models exhibit heterogeneous performance across the six task categories. Performance is highest in markets, reputation, and labor (NDS = 0.533) and lowest in networks, communication, and collective intelligence (NDS = 0.011). From the perspective of the knowledge-effect gap, the largest gap is observed in networks, communication, and collective intelligence (0.586), whereas the smallest gap appears in social psychology and group processes (0.013). These results suggest that the underlying mechanisms differ across categories. Some categories are relatively easy for the models to simulate, whereas networks, communication, and collective intelligence appear to be the most difficult.

\begin{table}[t]
\centering\small
\setlength{\tabcolsep}{3pt}
\caption{Directional performance by social science category. Category names are abbreviated. $n$ is the number of tasks in the category. Knowledge judgment and effect NDS are category-level proportions; K-E is the knowledge-effect gap.}
\label{tab:category-results}
\begin{tabular*}{\linewidth}{@{\extracolsep{\fill}}lrrrr@{}}
\toprule
Category & $n$ & Knowledge NDS $\uparrow$ & Effect NDS $\uparrow$ & K-E \\
\midrule
Behavioral econ. & 7 & 0.839 & 0.300 & 0.539 \\
Political econ. & 4 & 0.750 & 0.244 & 0.506 \\
Social psych. & 6 & 0.521 & 0.508 & 0.013 \\
Networks/comm. & 7 & 0.596 & 0.011 & 0.586 \\
Markets/rep. & 3 & 0.992 & 0.533 & 0.458 \\
Culture/norms & 3 & 0.717 & 0.150 & 0.567 \\
\bottomrule
\end{tabular*}
\end{table}

\paragraph{ANOVA}
We analyze all 1,200 task-seed simulation runs generated by the eight models to identify the main sources of variance across models. We conduct an ANOVA on the factors affecting simulation outcomes, using $\operatorname{same}$ and $\operatorname{std\_effect}$ as dependent variables. See Table~\ref{tab:anova}.

\begin{table}[t]
\centering\small
\setlength{\tabcolsep}{3pt}
\caption{Model, seed, and task effects on simulation outcomes.}
\label{tab:anova}
\begin{tabular*}{\linewidth}{@{\extracolsep{\fill}}llrrr@{}}
\toprule
Outcome & Effect & $F$ & $p$ & $\eta_p^2$ \\
\midrule
\textit{same} & model & 2.400 & ** & 0.015 \\
 & seed & 1.023 & ns & 0.004 \\
 & model $\times$ seed & 0.854 & ns & 0.021 \\
 & task & 21.248 & *** & 0.353 \\
\textit{std\_effect} & model & 3.414 & *** & 0.021 \\
 & seed & 0.948 & ns & 0.003 \\
 & model $\times$ seed & 0.838 & ns & 0.020 \\
 & task & 30.682 & *** & 0.440 \\
\bottomrule
\end{tabular*}
\end{table}

The results show that the effects of model and task are significant for both outcomes (all $p<0.05$), indicating that they are the primary sources of variance in simulation effects. Model performance varies across tasks, and systematic differences also exist across models. In contrast, seed and model $\times$ seed are not significant (all $p>0.05$), suggesting that simulation outcomes do not exhibit systematic differences in overall levels across different agent populations. This finding indicates that the models may not sufficiently capture the effects of within-human heterogeneity on outcomes.

In summary, the core factor behind model effect gaps is the task: task explains far more variance ($\eta_p^2=0.353$ and $0.440$) than model ($0.015$ and $0.021$), whereas no significant seed effect was detected.

\begin{table*}[t]
	\centering\small
	\caption{Causal chain length and simulation effects: overall and by knowledge judgment correctness. Parentheses contain task-clustered robust standard errors. Model FE and seed FE are controlled in all specifications.}
	\label{tab:causal-chain}
	\begin{tabular*}{\textwidth}{@{\extracolsep{\fill}}lrrrrrr@{}}
		\toprule
		Variable & \textit{same} & \textit{std\_effect} & knows=0 \textit{same} & knows=1 \textit{same} & knows=0 \textit{std\_effect} & knows=1 \textit{std\_effect} \\
		\midrule
		\textit{CausalChain} & $-0.205^{***}$ (0.049) & $-0.514^{***}$ (0.182) & 0.078 (0.100) & $-0.258^{***}$ (0.056) & 0.045 (0.241) & $-0.720^{***}$ (0.204) \\
		Constant & $-4.610$ (11.477) & $-1.514$ (4.795) & $-2.341$ (1.286) & 0.091 (1.053) & $-14.473^{***}$ (2.389) & 0.557 (3.035) \\
		Model FE & Yes & Yes & Yes & Yes & Yes & Yes \\
		Seed FE & Yes & Yes & Yes & Yes & Yes & Yes \\
		Controls & Yes & Yes & Yes & Yes & Yes & Yes \\
		$N$ & 1,200 & 1,200 & 190 & 1,010 & 190 & 1,010 \\
		Pseudo $R^2$ / $R^2$ & 0.172 & 0.250 & 0.458 & 0.222 & 0.587 & 0.265 \\
		\bottomrule
	\end{tabular*}
\end{table*}

\subsection{\emph{RQ4}: Do Model Results Corroborate the Bias Cascade Theory?}
We next examine the relationship between causal chain length in the task structure and the consistency of simulation outcomes. We define \textit{CausalChain} as a measure of the reasoning difficulty faced by the models. Tasks are coded into four levels according to the number of process stages required to proceed from the treatment to the primary outcome:
\begin{enumerate}[leftmargin=*]
\item \textbf{Level 1: Direct resolution.} The treatment directly changes the current observation or payoff, and the outcome is resolved through a single action.
\item \textbf{Level 2: One-step response or environmental resolution.} The treatment must pass through one response from another agent or one environmental resolution before reaching the primary outcome.
\item \textbf{Level 3: Cross-round aggregation of group states.} The outcome depends on the accumulation of group states over multiple rounds of interaction.
\item \textbf{Level 4: Strategic expectations, protocol aggregation, or cumulative cultural transmission.} The outcome requires strategic expectations, content aggregation, or accumulation across generations.
\end{enumerate}

The coding uses only the task specification, namely the manipulation and the primary outcome, and does not depend on the simulation results.

\begin{figure}[t]
    \centering
    \includegraphics[width=\linewidth]{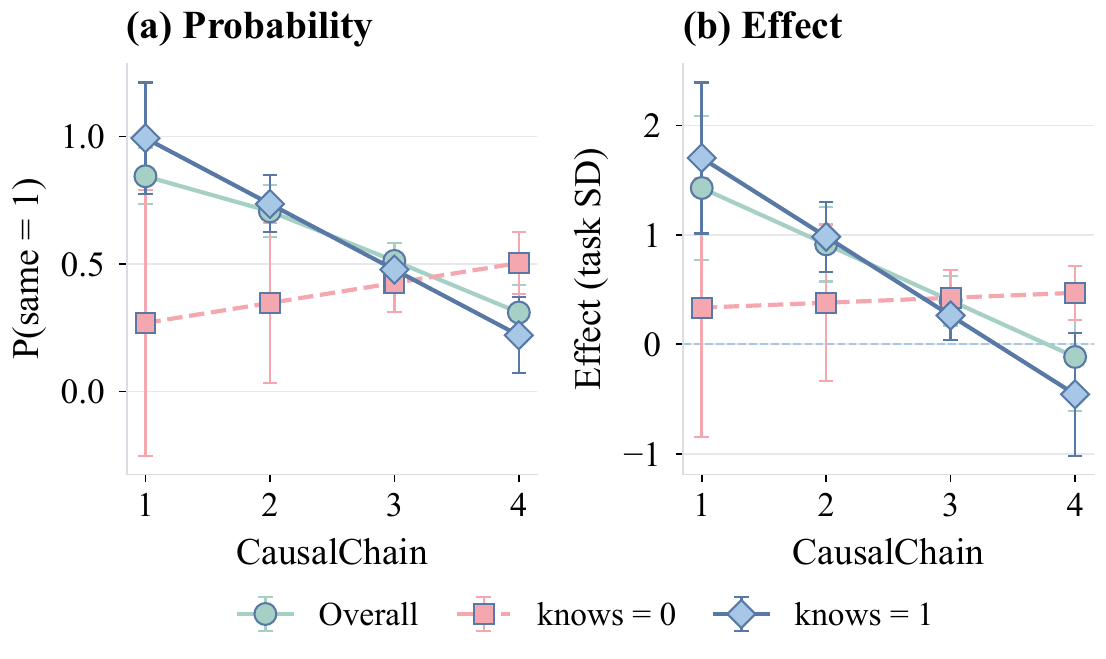}
    \caption{Adjusted predictions across causal chain levels. The left panel shows predicted same-direction probabilities; the right panel shows predicted standardized treatment effects (\textit{std\_effect}). Circles and solid lines denote the overall sample, squares and dashed lines denote knows=0 (incorrect or uncertain), and diamonds and solid lines denote knows=1 (correct). Error bars are 95\% confidence intervals.}
    \label{fig:regress-margins}
\end{figure}

For task $s$, agent seed $r$, and model $k$, we use $\operatorname{same}$ and $\operatorname{std\_effect}$ as dependent variables. The linear specification is
\begin{equation}
y_{s,r,k}=\alpha+\beta\,\mathrm{CausalChain}_s+\mathbf{X}_{s,r,k}'\gamma+\mu_k+\delta_r+\varepsilon_{s,r,k},
\end{equation}
where $\mathbf{X}_{s,r,k}$ denotes the control variables, and $\mu_k$ and $\delta_r$ denote model and seed fixed effects, respectively. Standard errors are clustered by task $s$. When $y=\mathrm{std\_effect}$, we use a linear fixed-effects estimator. When $y=\mathrm{same}$, we use a logit model with an analogous specification. Overall, causal chain length is significantly negatively associated with both the binary same-direction rate ($-0.205$, $p<0.01$) and the standardized within-task effect ($-0.514$, $p<0.01$). This indicates that as causal chain length increases, it becomes more difficult for the simulated effects to align with the reference direction. Table~\ref{tab:causal-chain} reports the full estimates, and Figure~\ref{fig:regress-margins} shows the adjusted predictions.

We also conduct separate regressions based on whether the knowledge judgment is accurate. Here, $\mathrm{knows}=1$ indicates that the model correctly identifies the reference direction in the knowledge judgment, whereas $\mathrm{knows}=0$ indicates an incorrect or uncertain judgment. After stratification, the negative relationship appears only in runs with correct knowledge judgments. When $\mathrm{knows}=1$, the coefficients are $-0.258$ ($p<0.01$) for \textit{same} and $-0.720$ ($p<0.01$) for \textit{std\_effect}. When $\mathrm{knows}=0$, the corresponding coefficients are 0.078 and 0.045, neither of which is statistically significant. The overall Pseudo $R^2$ / $R^2$ values are also higher than those of the full-sample specification, indicating that this stratification isolates a purer effect. These results show that models with accurate knowledge judgments are more susceptible to longer causal chains, providing further evidence for our claim that bias cascades lead to degraded simulation performance.

Taken together, these results corroborate the bias cascade theory: using \textit{CausalChain} as the key variable, longer causal chains are significantly associated with worse effect replication.

\section{Conclusion}
A metaphor succinctly captures our findings: 

\begin{quote}
	\textit{Coleman's boat cannot navigate the bias cascade.}
\end{quote}

Building on Coleman's boat, we identify two gaps between micro-level alignment and macro-level effect fidelity. First, aligned states may lead to different action rules. Second, matched marginal actions may generate different joint interactions and macro-level effects. Our proposed \emph{REFLEX} evaluates these gaps on 30 minimally intervened counterfactual tasks by comparing model-estimated average treatment effects with empirical findings and reporting a unified, single-model net directional score. Even the best-performing model, GLM-5.3, achieves scores of 0.847 on knowledge judgment and 0.453 on effect simulation, yielding a substantial gap of 0.393. Wording ablations further reveal high sensitivity to prompt phrasing, while regression analyses confirm that longer causal chains exacerbate performance degradation.

\bibliographystyle{ACM-Reference-Format}
\bibliography{ref}

\appendix
\section{Ethical Statement}\label{app:ethics}
This study did not recruit human participants and involved no real payments, registrations, contact with real-world individuals, or emergency-service calls. Helping, donating, and organizational participation were simulated actions.

\section{Limitations}\label{app:limitations}

First, each task uses only the direction of the effect reported in a single classic study as its reference point. For computational tractability, the platform resolves structured numerical outcomes using a rule-based engine. Although this approach improves consistency and reproducibility, it necessarily abstracts away from the complexity of real-world social institutions and interactions and may not capture the stability or generalizability of the underlying social mechanism across broader contexts.
Second, the population profiles are sampled from a publicly available dataset and consist primarily of adults within a specific age and education range. This design is necessary to ensure experimental comparability. However, these profiles may differ from the participants in the original studies and therefore may not fully represent the heterogeneity of those study populations.
Third, because the original studies provide limited data on participants' underlying thoughts, we cannot directly evaluate the theory by isolating the independent contribution of each step in Coleman's boat and comparing the resulting predictions with real-world observations. Instead, we assess this theoretical question indirectly through the length of the causal chain and examine the underlying mechanism in greater detail through case analyses.
Fourth, in diagnosing the bias cascade, we manually code causal-chain length on a scale from 1 to 4 to quantify its influence. This coding procedure may introduce a degree of subjectivity.

\section{Terminology}\label{app:terminology}
This paper involves four related terms that operate at different levels.
\emph{Knowledge judgment} and \emph{effect simulation} are two experimental settings at the measurement level: the former asks a model to predict, independently of the simulation environment and using only the task background, the direction of the effect induced by a given social factor; the latter places LLM agents in a constructed social environment, where they act on locally available observations, and aggregates their individual actions into a macro-level outcome from which the treatment effect is estimated.
\emph{Knowledge alignment} and \emph{effect replication} are the target properties corresponding to these two settings, at the construct level: knowledge alignment refers to the agreement between a model's knowledge of social situations and their consequences and human knowledge, whereas effect replication refers to the agreement between the direction of the treatment effect and the conclusions of human studies.

\section{Core Motivation}\label{app:motivation}

\subsection{The Importance of Causal Modeling}

Recent years have seen substantial progress in LLM-based social simulation and its evaluation in the social sciences. Our study differs from this line of work by placing causal interventions at the center of simulation design and using the resulting treatment effects as the primary criterion for effect replication.

Prior work has introduced end-to-end benchmarks for evaluating simulation capabilities \cite{hu2026simbench,liu2026humanstudy}, but these benchmarks do not primarily assess effect replication. Instead, they generally characterize model capabilities through direct comparisons between simulated and observed responses or outcomes. This capability-oriented evaluation paradigm has two main limitations. Existing frameworks often assess associational agreement without incorporating causal designs. Although numerous evaluation metrics have been proposed, these metrics may not capture the mechanisms central to social simulation. Optimizing against them may reward spurious correlations while obscuring the sources of simulation failure. A causal approach to social simulation instead requires an explicit social environment in which treatment effects can be examined through counterfactual social experiments.

These considerations motivate a counterfactual social experiment benchmark centered on cross-level causal processes in social simulation. From a causal perspective, average treatment effects, particularly their directions, provide a principled basis for assessing whether simulations reproduce population-level intervention responses. This is the core motivation behind \emph{REFLEX}.

\subsection{Coleman's Boat as Explanatory Framework}

Existing research on social simulation has also made limited use of classical social science theory. Drawing on Coleman's boat, our study identifies a gap between knowledge alignment and effect replication, and develops a theoretical account of this gap.

Coleman's boat is a sociological framework that explains how macro-level social conditions produce collective outcomes through micro-level actions and interactions. Its central premise is that empirical associations between macro-level variables are insufficient for a sociological explanation. An explanation must also identify how social conditions shape individual actions and how those actions, in turn, generate macro-level outcomes. The framework thus connects macro-level social conditions, micro-level states, individual actions, and collective outcomes through situational, action-formation, and transformation mechanisms. In our study, it provides a unified analytical framework for considering both structural constraints and individual agency. This perspective calls for modeling actors who adapt their behavior within dynamic environments, rather than restricting evaluation to the static, one-shot choices examined in much of the existing literature.

Some studies have considered the distinction between knowing and acting \cite{jang2026instruction}. However, most operationalize social simulation through single-turn, questionnaire-style evaluations. Their primary focus remains agreement with human responses rather than the action capabilities required in dynamic settings. Such evaluations mainly assess responses associated with individual characteristics, while leaving interaction processes largely unexamined. These evaluations remain appropriate for psychological research questions centered on individual responses. However, many social phenomena cannot be understood solely by aggregating individually aligned responses. They also depend on how knowledge judgments translate into actions and how those actions interact over time.

This perspective informs the design of \emph{REFLEX}, which evaluates model capabilities through outcomes generated in an explicit simulation environment. Coleman's boat further guides our analysis of the mechanisms that may account for the observed gap between knowledge alignment and effect replication.

\section{Methods and Assumptions}\label{app:assumptions}

\subsection{Model Assumptions}

Within Coleman's boat framework, macro-level knowledge of a social context $m$ is mapped through situational mechanisms to a distribution of micro-level states $Q_q(z \mid m)$, where $z$ represents individuals' circumstances and states. Our theoretical analysis introduces an idealized assumption about LLMs' understanding of social contexts: when a model's knowledge of a macro-level social condition is aligned with human knowledge, the simulation system can, under idealized conditions, map its representation of that condition to the corresponding human distribution of micro-level states. The rationale is that relevant social knowledge includes relationships between social conditions and individual circumstances and states, providing a basis for this macro-to-micro mapping. Our experiments provide evidence of knowledge alignment for the knowledge judgments we evaluate, although they do not directly establish alignment at the micro-level state.

We treat micro-level state alignment as an assumption for two reasons. First, we cannot fully reconstruct the social contexts and individual circumstances of the original experiments, and therefore lack verifiable micro-level ground truth. Second, even if those contexts could be reconstructed, the claim that AI systems and humans share identical subjective perceptions would remain difficult to operationalize and test. We therefore focus empirical evaluation on observable macro-level outcomes and adopt agreement in treatment-effect direction as a falsifiable criterion. Specifically, under corresponding counterfactual experimental manipulations, we test whether the direction of the model-estimated macro-level treatment effect agrees with that reported in prior empirical social science research.

We also examine the alternative explanation that the observed failures arise from limitations in environmental design. To this end, we verify that each environment admits a configuration capable of reproducing the empirical effect direction. This check assesses whether the environments preclude effect replication by design; it does not establish sufficiency in all respects. For aspects that cannot be resolved experimentally, we rely on theoretical analysis.

\subsection{Probing the Bias Cascade}

The bias-cascade account within Coleman's boat involves two core transmission stages: the mapping from micro-level states $Q$ to action-selection rules $\rho$, and the aggregation of marginal action distributions into macro-level outcomes $Y$ through interaction trajectories $\mathcal{T}$. Our experiments cannot directly identify these stages. First, Coleman's boat is an explanatory framework rather than a statistical measurement model. Second, the intermediate processes are not identifiable from the available data. Separately identifying the two stages would require complete individual-level trajectories under counterfactual conditions, together with the joint response distribution of the same individuals within an interaction network. Such data are not available for the empirical studies considered here. Direct identification of the intermediate processes underlying a bias cascade is therefore beyond the scope of the current design.

Accordingly, we use the bias cascade as an explanatory account of simulation failure, rather than claiming to directly identify its causal contribution. Nevertheless, our analysis extends beyond theoretical interpretation. To investigate the gap between accurate knowledge judgments and unsuccessful effect replication, we introduce the variable \textit{CausalChain} as an indirect empirical test of the bias-cascade account. This variable characterizes the causal and interaction structure linking individual actions to macro-level outcomes, allowing us to examine observable patterns predicted by the framework.

We find that effect replication rates decline as causal-chain length increases and interactions become more complex. These macro-level patterns are consistent with the framework's predictions and support the interpretation that simulation failures reflect the propagation and amplification of biases during simulation rather than deficiencies in static knowledge alone.

\subsection{Population and Sampling}

For all \emph{REFLEX} tasks, agent populations are sampled from SAPA, a shared, internally heterogeneous general-population pool. Varying the agent seed resamples agents from this fixed pool rather than reconstructing the demographic sampling frame of each original human study. Differences between our sample pool and the populations studied in the original experiments limit the validity and generalizability of our findings. At the same time, using a common population pool provides a consistent basis for cross-task comparisons and reduces variation attributable to task-specific sampling choices. Our design therefore prioritizes comparability across tasks over population matching to each original study.

We find no statistically significant effect of the agent seed under the current design. This result indicates that resampling within the SAPA pool does not produce detectable variation in the estimated treatment effects. It does not establish an absence of meaningful population heterogeneity: results may differ under sampling frames matched to the original experiments, and the observed stability may reflect limited variation in the sampling pool, limited model sensitivity to population characteristics, or both.

In particular, reduced heterogeneity in model-generated actions may itself contribute to the effect replication failures observed in this study. This explanation is compatible with the proposed theoretical framework, and our analysis does not rule it out. Because population heterogeneity is not the primary focus of our research question, we retain a common-pool design for cross-task evaluation and leave population-matched sampling and the explicit analysis of heterogeneity to future work.

\section{Bootstrap Confidence Interval Calculation}\label{app:bootstrap}
For the primary analysis, we resample $S$ tasks with replacement and retain all valid seeds for each sampled task; seeds are not resampled. Let $s_j^{*(b)}$ denote the task drawn for position $j$ in bootstrap replicate $b$, and let $x_j^{*(b)}=x_{s_j^{*(b)}}$ be its task score. The bootstrap estimate is
\begin{equation}
\widehat{\theta}^{*(b)}=\frac{1}{S}\sum_{j=1}^{S}x_j^{*(b)}.
\end{equation}
We use $B=2{,}000$ replicates, random seed 555, and bias-corrected and accelerated (BCa) intervals.

\section{Rule-Based Baseline}\label{app:baseline}
\subsection{Evaluation Objective and Setup}

One might object to the thesis of this paper: model performance is not determined directly by deficiencies in LLM social simulation capability; instead, the difference in differences is driven by internal simulation noise, that is, the poor measured model performance arises because the system itself admits no reasonable feasible solution. We rule out this possibility using a rule-based baseline.

An alternative explanation for the performance decline observed in the simulation experiments is not that the models lack the required capabilities, but that the task design makes high performance fundamentally unattainable. This section evaluates and rules out this possibility.

We examine whether the directional effects specified by \emph{REFLEX} are attainable in the current task environments. If an explicit rule-based model using only local observations can produce population-level effects that match the reference under the same environment, action constraints, and settlement rules, then the tasks are directionally satisfiable. To this end, we implement a purely rule-based agent model in accordance with the large language model experimental setup. All configurations are generated once using the default heuristic rules, without additional validation or hyperparameter tuning.

We prompt GLM-5.3 to directly generate a rule that conforms to real-scenario variation. Its information about how the environment is executed is omniscient, but the rule is generated by the model in a single pass, without repeated tuning to improve the score. Notably, this model may not produce the best possible rule; the values obtained on this basis should therefore represent a theoretical lower bound on the highest achievable score for the problems in the scenario.

\subsection{Metric Results}

The evaluation uses \emph{REFLEX}'s 30 environments, action space, primary metric, and prescribed directions, with the environment seed fixed at 42. Each task is evaluated once for each subject seed from 1 to 5. The rules receive only observations available to the subject and do not access hidden global information. Actions continue to be validated and settled according to the original \emph{REFLEX} rules.

The resulting NDS is 0.853, with an average of 27.4 tasks per seed exhibiting the prescribed direction, see Table~\ref{tab:rule-baseline}. This result substantially exceeds the best NDS of 0.453 obtained in the current simulation experiments.

Moreover, after averaging across the five subject seeds for each task, the estimated effect direction matches the human reference for all 30 tasks. Thus, an explicit rule-based simulator constructed for the current task set can reproduce the prescribed directions across all 30 tasks under the default environment seed, although it does not do so for every individual subject seed.

These results show that the reference directions in \emph{REFLEX} are attainable rather than inherently unsatisfiable. The LLM's low NDS is therefore more plausibly explained by limitations in action formation and interaction sequencing than by a benchmark design that excludes the correct direction.

These experiments rule out the hypothesis that the poor measured model performance is because the system itself has no reasonable feasible solution, and they support the hypothesis proposed in this paper.

\begin{table}[ht]
\centering\small
\setlength{\tabcolsep}{1.5pt}
	\caption{Rule-based action baseline results. Counts are mean numbers of tasks per subject seed; DAR, DRR, and NDS use the 150 task-by-seed units.}
\label{tab:rule-baseline}
\begin{tabular*}{\linewidth}{@{\extracolsep{\fill}}lrrrr@{}}
\toprule
Configuration & + / 0 / - & DAR $\uparrow$ & DRR $\downarrow$ & NDS $\uparrow$\\
\midrule
Rule-based & \same{27.4}/\zero{0.8}/\opp{1.8} & 0.913 & 0.060 & 0.853\\
\bottomrule
\end{tabular*}
\end{table}

\section{Survey-Calibrated Profiles}\label{app:profiles}
The survey-calibrated mode uses the public SAPA sample. It contains 25 Big Five self-report items, five for each of A, C, E, N, and O, on a 1-6 Likert scale, together with self-reported age and education coded from 1 to 5. The raw snapshot has 2,800 records. Keeping records with valid answers to all personality items, age 18-100, and education 1-5 yields 2,185 profiles; Table~\ref{tab:survey-profiles} reports their descriptive statistics.

\begin{table}[ht]
\centering\small
\caption{Descriptive statistics for survey-calibrated profiles ($N=2{,}185$).}
\label{tab:survey-profiles}
\begin{tabular*}{\linewidth}{@{\extracolsep{\fill}}lr@{}}
\toprule
Variable & Mean (SD)\\
\midrule
Age in years & 30.3 (10.5)\\
Education, 1-5 & 3.22 (1.09)\\
Agreeableness, 1-6 & 4.69 (0.89)\\
Conscientiousness, 1-6 & 4.32 (0.95)\\
Extraversion, 1-6 & 4.15 (1.07)\\
Emotional stability, 1-6 & 3.86 (1.20)\\
Openness, 1-6 & 4.62 (0.80)\\
\bottomrule
\end{tabular*}
\end{table}

\section{Simulation Platform Architecture}\label{app:platform}
\begin{figure}[t]
    \centering
    \includegraphics[width=\linewidth]{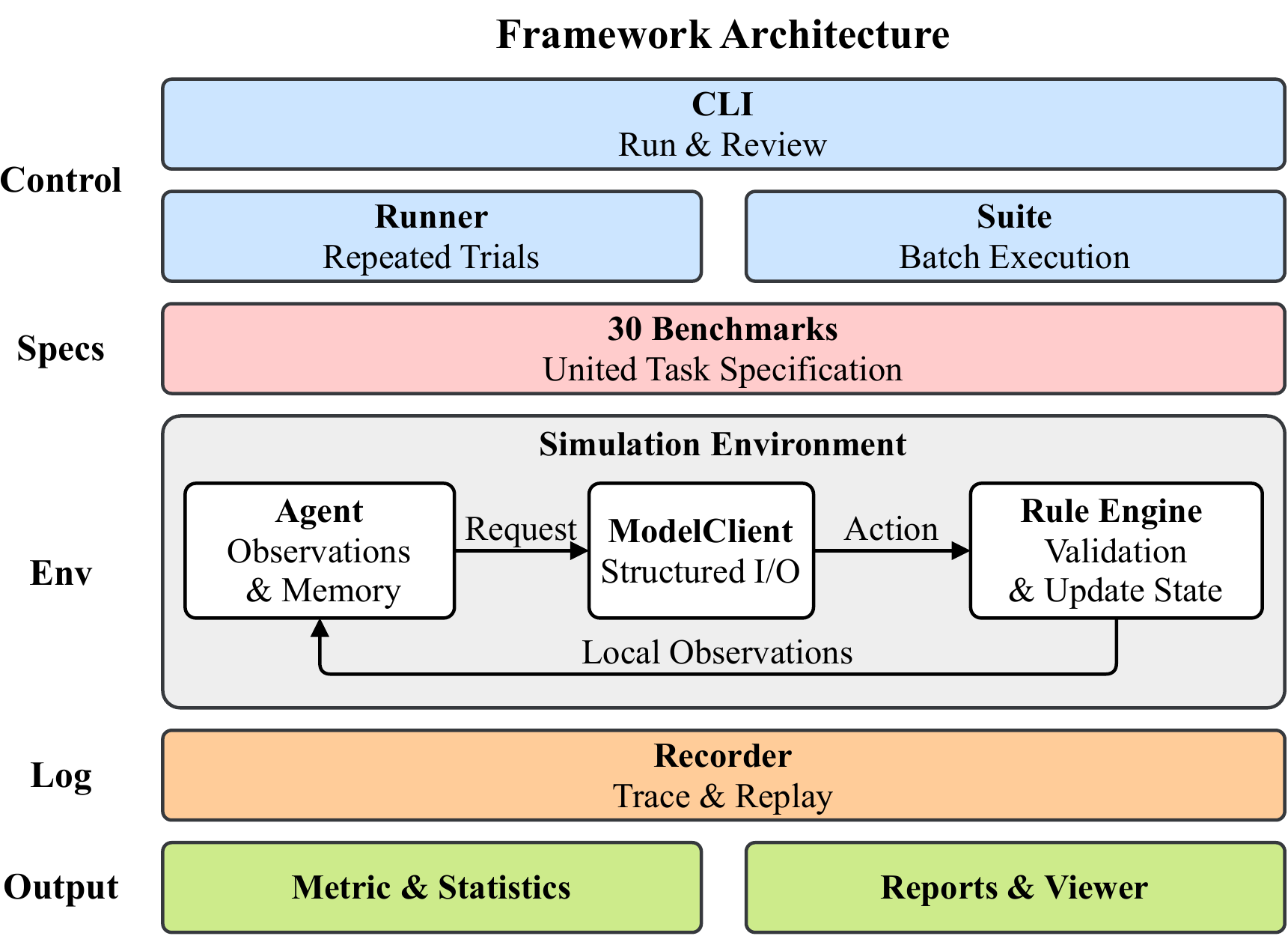}
    \caption{Layered architecture of our framework, which separates task specification, trial execution, the rule-based simulation environment, trace logging, and metric reporting.}
    \label{fig:architecture}
\end{figure}

\subsection{Layered Responsibilities}
The layered architecture of our framework is shown in Figure~\ref{fig:architecture}. The platform consists of layers for task specifications, the environment, populations and agents, logging, and result generation. The responsibilities of these layers are strictly separated: the environment generates the information visible to agents and adjudicates the outcomes of actions, the Python code executes the rules and updates the state, and the logging layer stores the complete event chain.

The platform follows a unified execution procedure:
\begin{enumerate}[leftmargin=*]
\item \textbf{Generate observations.} Based on the current global state, the environment generates a locally visible observation for each agent.
\item \textbf{Request decisions.} The system provides the model with the task constraints, the agent's own information, and the observation for the current round, and requests a structured action.
\item \textbf{Validate responses.} The returned result is first checked for format validity and value ranges. If the response is invalid, the system issues a retry under the same decision ID. The entire retry process is recorded.
\item \textbf{Execute rules.} Based on the accepted actions, the Python rule engine uniformly computes outcomes including payoffs, resource changes, network updates, message delivery, and market settlement.
\item \textbf{Record transitions.} The system writes the agents' actions and the newly settled state to the event stream and updates the run record.
\item \textbf{Proceed to the next round.} The observations for the next round are always generated from the state after the previous round has been fully settled.
\end{enumerate}

Thus, models can only propose actions and cannot directly modify the environment state. All effects on the environment must be validated and settled by the rule engine.

\subsection{Information Isolation and Recomputability}
The ground truth required for evaluation is not exposed to agents. The runtime environment projects each agent's observable information based on its role, location, and network relationships. Agents cannot access other agents' private information, hidden utilities, undelivered messages, or evaluation answers. Agents may submit only valid actions and cannot directly modify resources, relationships, matching outcomes, or market states.

All state changes are executed by Python code, including resource transfers, network relationship updates, agent matching, market clearing, and penalties for violations. This design separates model decisions from the actual consequences imposed by the environment and prevents models from bypassing the experimental rules through their outputs.

\section{Mechanistic Analysis of Examples}\label{app:mechanistic}
\subsection{Positive: Punishment Raises the Cost of Low Contribution}

In the public-goods task, costly punishment is expected to increase final-round contributions. Participants allocate fixed resources to a public good, whose total contribution is multiplied and redistributed. Punishment costs the sender and reduces the recipient's payoff. After observing contributions, agents decide whether to punish, and the resulting payoffs affect the next round.

The decision rule is that punishment minus no punishment is greater than zero. A positive effect indicates that the model understands punishment as a way to deter free riding, uses it conditionally, and causes recipients to increase later contributions.

For seed 1, contribution rates under no punishment are 0.4625, 0.2875, 0.1875, 0.0875, and 0, compared with 0.4875, 0.5625, 0.5625, 0.5625, and 0.5625 under punishment. The model uses punishment, predicts the correct direction, and produces higher subsequent contributions. This shows positive replication in knowledge, conditional action, and interaction continuation.

After the initial punishment, the punishment group maintains higher contributions without further punishment. Across repeated runs, punishment is used in every run and the final-round contribution difference is always positive.

\paragraph{Reflection}
This is a fully correct simulation result, and it shows that the model is capable of handling some situations. Overall, the model's behavior here is clear and correct, which may be related to the fact that the causal chain length of this task is only 2.

\subsection{Null: Universal Upvoting Erases the Initial-Rating Difference}

The initial-rating task tests whether changing a comment's initial score from 0 to $+1$ increases later upvotes. Readers sequentially upvote, downvote, or abstain. Only the displayed initial score differs across conditions, and each reader observes the accumulated score.

The decision rule is that initial-upvote minus neutral is greater than zero. The comment concerns extending library hours during exam periods. Each reader observes the current score and updates it for the next reader. The task also includes an initial-downvote condition.

The model knows that an initial upvote should increase later upvotes, but all readers upvote under every condition. In Table~\ref{tab:voting-outcomes}, subsequent upvote rates are 1.0000 under neutral, initial-upvote, and initial-downvote conditions. Final scores are 12, 13, and 11, respectively. Thus, the initial signal is represented in the input but does not affect behavior. The model shows directional knowledge but fails to reproduce human condition sensitivity.

\begin{table}[t]
\centering\small
\caption{Seed 1 voting outcomes under the three initial-rating conditions.}
\label{tab:voting-outcomes}
\setlength{\tabcolsep}{4pt}
\begin{tabular*}{\linewidth}{@{\extracolsep{\fill}}lrrrr@{}}
\toprule
Condition & Initial score & \shortstack{Votes by\\12 readers} & \shortstack{Subsequent\\upvote rate} & Final score \\
\midrule
Neutral & 0 & All upvoted & 1.0000 & 12 \\
Initial upvote & $+1$ & All upvoted & 1.0000 & 13 \\
Initial downvote & $-1$ & All upvoted & 1.0000 & 11 \\
\bottomrule
\end{tabular*}
\end{table}

Repeated runs produce the same result. The initial vote changes the displayed score, but every subsequent reader chooses the same action. The null effect therefore reflects identical action distributions rather than genuine condition sensitivity.

\paragraph{Reflection}
The model upvotes in essentially every case, whereas a situation in which everyone upvotes is almost nonexistent in reality. This is a bias in the mapping from states to actions. The reason may be that training overfits the model to user-pleasing behavior: although the model knows that real people may not upvote, it still upvotes out of a habit acquired in user interaction.

\subsection{Reversed: Local Adoption Does Not Become Sustained Diffusion}

The network task tests whether clustered networks produce more adoption than random networks. The manipulation changes only network connections. Each participant has four contacts, and nodes 6 and 9 initially adopt.

For seed 1, cumulative new adopters are 3, 7, 7, and 7 in the random network, compared with 4, 4, 4, and 4 in the clustered network. Fourteen non-initial members are notified in the random network and eight in the clustered network. Adoption among notified members is 0.5000 in both networks, but final-round adoption rates among all 14 non-initial members are 0.5000 and 0.2857, respectively.

The outcome is the final-round adoption rate among all 14 non-initial members. The decision rule is that clustered minus random is greater than zero. Clustered networks provide more overlapping social reinforcement, which should support adoption when multiple sources are required.

\begin{table}[t]
\centering\small
\caption{Seed 1 cumulative diffusion trajectory and exposure in the random and clustered networks.}
\label{tab:diffusion-trajectory}
\begin{tabular*}{\linewidth}{@{\extracolsep{\fill}}lrr@{}}
\toprule
Measure & Random & Clustered \\
\midrule
New adopters, round 1 & 3 & 4 \\
New adopters, round 2 & 7 & 4 \\
New adopters, round 3 & 7 & 4 \\
New adopters, round 4 & 7 & 4 \\
Members ever notified & 14 & 8 \\
Adoption among notified & 7/14 = 0.5000 & 4/8 = 0.5000 \\
Final-round adoption (all 14) & 7/14 = 0.5000 & 4/14 = 0.2857 \\
\bottomrule
\end{tabular*}
\end{table}

The clustered network leads initially, but peripheral agents reject adoption, preventing further exposure. The random network reaches more dispersed contacts. Across five seeds, the effect is aligned once, null once, and reversed three times, with a mean of $-0.14286$. This negative result is specific to the model's behavioral response and does not imply that random networks are generally superior.

In seed 1, the initial adopters and first-round adopters in the clustered network occupy the same local area. Their combined contacts reach only eight non-initial members. Four bridge nodes connect this area to the rest of the network, but they do not adopt, so additional nodes receive no notifications. In the random network, adopters' contacts are more dispersed, allowing all 14 non-initial members to be notified.

Both networks have the same adoption rate among notified participants. The negative effect therefore results mainly from limited exposure rather than lower conversion. Using only notified participants as the denominator yields no difference, whereas using all 14 captures both exposure and conversion. Across seeds, clustering does not produce a stable advantage, and the average effect is reversed.

\paragraph{Reflection}
This is the most typical example, and it shows that the causes of the model's results can be complex. Two effects are actually at work here. First, the adoption rate does not in fact increase as the network becomes more tightly connected, whereas real people may be more likely to adopt when they have more contacts; this is a bias in the mapping from states to actions. Second, deeper amplification effects arise during interaction: for example, individuals at key positions in the network may block the process, which is the opposite of what happens in many real situations; this is a bias generated in the interaction process.

\section{Three Representative Trajectories}\label{app:trajectories}
\begin{figure}[t]
    \centering
    \includegraphics[width=\linewidth]{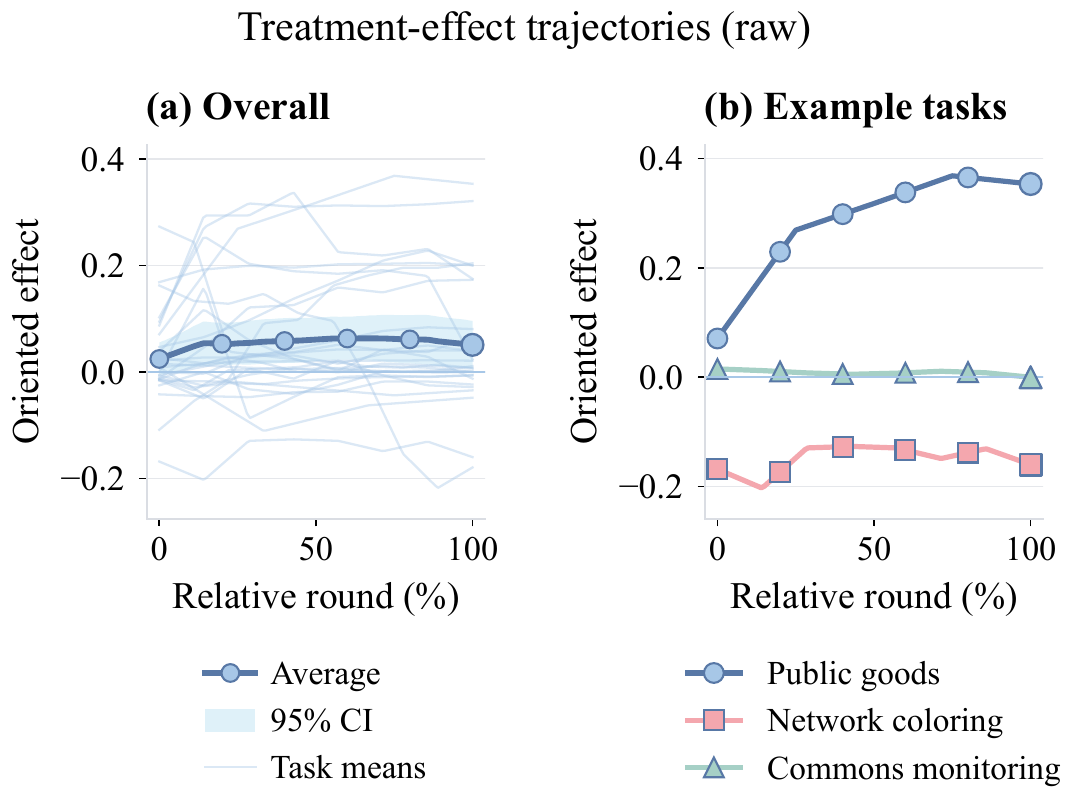}
    \caption{Representative treatment-effect trajectories in raw units (left: 28 tasks; right: three task means).}
    \label{fig:trajectory-raw}
\end{figure}

\begin{figure}[t]
    \centering
    \includegraphics[width=\linewidth]{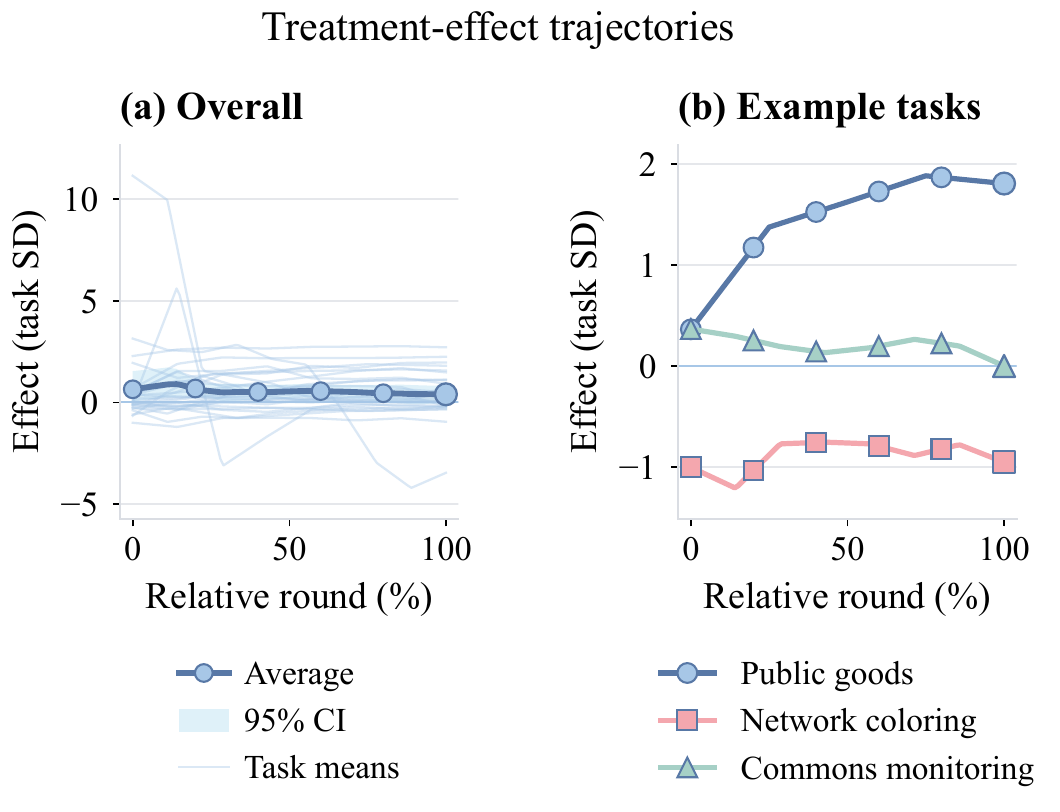}
    \caption{Same trajectories as Figure~\ref{fig:trajectory-raw}, in task-standard-deviation units.}
    \label{fig:trajectory-standardized}
\end{figure}

Regression coefficients summarize task-level associations but do not show how treatment effects develop across rounds. Figure~\ref{fig:trajectory-raw} presents three representative trajectories on a common relative-round axis: a persistent positive effect for public-goods punishment, a persistent negative effect for network coloring, and a near-zero effect for commons monitoring. Each curve averages across eight models and five subject seeds, with equal weights assigned to tasks. The left panel shows all 28 synchronous-round tasks, their equal-weight mean, and 95\% bootstrap confidence intervals. Figure~\ref{fig:trajectory-standardized} reports the same trajectories in \textit{std\_effect}; the numerical discussion below uses the raw differences in Figure~\ref{fig:trajectory-raw}.

Public-goods punishment compares a punishment condition with a no-punishment condition, using the final contribution rate as the primary outcome. The treatment advantage increases from about 0.07 in the first round to about 0.37 by round 4, then remains near 0.35 in the final round. All five subject seeds and eight models produce positive final-round effects. Thus, the result is not limited to a single round. Punishment is followed by higher contributions from the target, and the difference persists over time. This is a stable and sustained positive trajectory in this study.

Network coloring compares a rewired network with long-range links against a regular grid. The outcome is the share of edges connecting nodes with different colors. The rewired network does not produce the expected coordination advantage. The trajectory remains negative, at about $-0.17$ in the first round, $-0.20$ in round 2, between approximately $-0.15$ and $-0.13$ thereafter, and about $-0.16$ in the final round. All five seeds and eight models produce negative effects, and about 88\% of the 40 runs are negative. This is a stable negative trajectory.

Commons monitoring compares visible individual extraction with aggregate-only information, using the resource-stock rate as the primary outcome. The two conditions produce nearly identical trajectories, with treatment differences between 0 and 0.015 and a final-round difference of zero. This does not provide clear evidence of a null simulation effect. In both conditions, approximately 68 to 69 units are removed in the first round, reducing the stock from 80 to 26-28 units. A further 25 to 26 units are removed in the second round, leaving only 4 to 5\% of the stock, and more than 80\% of runs reach zero. Later rounds remain near this lower bound, with both conditions removing about three units per round.

\section{Causal Chain Coding}\label{app:causal-coding}
All 30 tasks are assigned a causal chain level using the four-level rubric in Section~\ref{sec:results}. The coding is task-level: it uses only the registered manipulation and the primary outcome, not the simulated effect directions or model outputs. The resulting distribution is 4 tasks at Level 1, 9 tasks at Level 2, 9 tasks at Level 3, and 8 tasks at Level 4 (mean $=2.70$, SD $=1.00$). The tasks assigned to each level are listed below.

\begin{table}[t]
\centering
\caption{Causal chain levels and the tasks assigned to each level. The Count column gives the number of tasks at each level.}\label{tab:causal-coding}
\small
\begin{tabular}{@{}p{0.08\linewidth}p{0.08\linewidth}p{0.69\linewidth}@{}}
\toprule
Level & Count & Tasks\\
\midrule
1 & 4 & Initial ratings and following, Repeated estimation and social information, Multi-round signals and action cascades, Weak-link coordination and group size.\\
2 & 9 & Public goods and punishment, Rotating help and public reputation, Voluntary entry to public projects, Peer rewards after contribution, Market-price feedback and convergence, Stable identity in labor markets, Fulfillment reputation in online trade, Cumulative mitigation and collective risk, Intergroup competition and coordination.\\
3 & 9 & Wealth visibility, Social reinforcement and diffusion, Dynamic partner network, Communication in renewable commons, Information levels in commons extraction, Interaction networks and conventions, Committed minorities and conventions, Centralized and decentralized belief updating, Cultural-market popularity accumulation.\\
4 & 8 & Hidden-information puzzle, Gossip and temporary exclusion, Membership migration, Network coloring with local information, Collaborative search and sharing, Intermittent contact and collective problem solving, Group recombination and cumulative design, Communication organization and team aggregation.\\
\bottomrule
\end{tabular}
\end{table}

\section{Tasks and Comparisons}\label{app:tasks}

Table~\ref{tab:task-registry} summarizes each task by its core comparison, expected direction, original study, and task-level results. All 30 tasks are supported by stable bibliographic sources. Twenty-seven tasks have a positive expected effect, while three have a negative expected effect. Each manipulation varies only its registered social factor, such as information visibility, an institutional switch, network structure, or risk level.

Table~\ref{tab:task-registry} also reports task-level metrics, revealing substantial variation in performance. Strong positive effects arise in tasks with salient and locally actionable manipulations, including Public goods and punishment with an NDS of 0.900, Repeated estimation and social information with a value of 1.000, Intermittent contact and collective problem solving with a value of 1.000, and Weak-link coordination and group size with a value of 1.000. In contrast, performance is poor when success depends on aggregating dispersed information or sustaining coordinated exploration, as observed in Voluntary entry to public projects with a value of $-0.800$, Network coloring with local information with a value of $-0.650$, and Collaborative search and sharing with a value of $-0.400$. Tasks with weak or ambiguous manipulations, such as Initial ratings and following and Cultural-market popularity accumulation, cluster near zero at 0.000. Overall, benchmark difficulty is highly task-dependent. Failures are concentrated in tasks that require information aggregation, coordinated exploration, or sustained diffusion, rather than simple local responses. This pattern is consistent with our categorical analysis and assessment of causal chain length.

\clearpage
\onecolumn
{\small
\begin{longtable}{@{}lp{0.21\linewidth}p{0.14\linewidth}p{0.18\linewidth}@{\hspace{4pt}}p{0.10\linewidth}rrr}
\caption{Task-level averages under the dynamic protocol, aggregated over five subject seeds and eight models. Citations in the Task column identify the source of each task.}\label{tab:task-registry}\\
\toprule
\# & Task & Default structure & Core comparison & + / 0 / - & DAR $\uparrow$ & DRR $\downarrow$ & NDS $\uparrow$\\
\midrule
\endfirsthead
\toprule
\# & Task & Default structure & Core comparison & + / 0 / - & DAR $\uparrow$ & DRR $\downarrow$ & NDS $\uparrow$\\
\midrule
\endhead
1 & Wealth visibility \citeapp{nishi2015inequality} & 8 agents $\times$ 5 rounds & visible $-$ invisible $<0$ & \same{1.5}/\zero{0.9}/\opp{2.6} & 0.300 & 0.525 & $-0.225$\\
2 & Hidden-information puzzle \citeapp{stasser1985pooling} & 4 agents $\times$ 3 rounds & shared $-$ unshared $>0$ & \same{2.5}/\zero{2.4}/\opp{0.1} & 0.500 & 0.025 & 0.475\\
3 & Public goods and punishment \citeapp{fehr2002altruistic} & 4 agents $\times$ 5 rounds & punishment $-$ none $>0$ & \same{4.5}/\zero{0.5}/\opp{0.0} & 0.900 & 0.000 & 0.900\\
4 & Social reinforcement and diffusion \citeapp{centola2010spread} & 16 agents $\times$ 4 rounds & clustered $-$ random $>0$ & \same{1.2}/\zero{1.8}/\opp{2.0} & 0.250 & 0.400 & $-0.150$\\
5 & Initial ratings and following \citeapp{muchnik2013social} & 12 sequential agents & upvote $-$ neutral $>0$ & \same{0.6}/\zero{3.8}/\opp{0.6} & 0.125 & 0.125 & 0.000\\
6 & Dynamic partner network \citeapp{rand2011dynamic} & 8 agents $\times$ 8 rounds & rewiring $-$ fixed $>0$ & \same{4.2}/\zero{0.4}/\opp{0.4} & 0.850 & 0.075 & 0.775\\
7 & Rotating help and public reputation \citeapp{wedekind2000cooperation} & 8 agents $\times$ 8 rounds & public $-$ private $>0$ & \same{2.0}/\zero{0.2}/\opp{2.8} & 0.400 & 0.550 & $-0.150$\\
8 & Gossip and temporary exclusion \citeapp{feinberg2014gossip} & 8 agents $\times$ 8 rounds & gossip $-$ private reports $>0$ & \same{3.4}/\zero{0.1}/\opp{1.5} & 0.675 & 0.300 & 0.375\\
9 & Membership migration \citeapp{gurerk2006competitive} & 8 agents $\times$ 8 rounds & choice $-$ assigned $>0$ & \same{2.2}/\zero{0.1}/\opp{2.6} & 0.450 & 0.525 & $-0.075$\\
10 & Voluntary entry to public projects \citeapp{semmann2003volunteering} & 8 agents $\times$ 8 rounds & optional $-$ mandatory $>0$ & \same{0.5}/\zero{0.0}/\opp{4.5} & 0.100 & 0.900 & $-0.800$\\
11 & Peer rewards after contribution \citeapp{rand2009positive} & 8 agents $\times$ 8 rounds & reward $-$ none $>0$ & \same{3.6}/\zero{0.1}/\opp{1.2} & 0.725 & 0.250 & 0.475\\
12 & Communication in renewable commons \citeapp{ostrom1992covenants} & 8 agents $\times$ 8 rounds & discussion $-$ notes $>0$ & \same{4.0}/\zero{0.2}/\opp{0.8} & 0.800 & 0.150 & 0.650\\
13 & Information levels in commons extraction \citeapp{sell1991levels} & 8 agents $\times$ 8 rounds & identified $-$ aggregate $>0$ & \same{2.1}/\zero{1.1}/\opp{1.8} & 0.425 & 0.350 & 0.075\\
14 & Network coloring with local information \citeapp{kearns2006experimental} & 12 agents $\times$ 8 rounds & rewired $-$ lattice $>0$ & \same{0.9}/\zero{0.0}/\opp{4.1} & 0.175 & 0.825 & $-0.650$\\
15 & Interaction networks and conventions \citeapp{centola2015spontaneous} & 12 agents $\times$ 10 rounds & mixing $-$ local $>0$ & \same{3.6}/\zero{0.0}/\opp{1.4} & 0.725 & 0.275 & 0.450\\
16 & Committed minorities and conventions \citeapp{centola2018experimental} & 12 agents $\times$ 10 rounds & large $-$ small minority $>0$ & \same{1.9}/\zero{2.0}/\opp{1.1} & 0.375 & 0.225 & 0.150\\
17 & Repeated estimation and social information \citeapp{lorenz2011social} & 8 agents $\times$ 8 rounds & social $-$ independent $<0$ & \same{5.0}/\zero{0.0}/\opp{0.0} & 1.000 & 0.000 & 1.000\\
18 & Centralized and decentralized belief updating \citeapp{becker2017network} & 8 agents $\times$ 8 rounds & decentralized $-$ centralized $>0$ & \same{4.6}/\zero{0.0}/\opp{0.4} & 0.925 & 0.075 & 0.850\\
19 & Collaborative search and sharing \citeapp{mason2012collaborative} & 12 agents $\times$ 10 rounds & random $-$ local $>0$ & \same{1.5}/\zero{0.0}/\opp{3.5} & 0.300 & 0.700 & $-0.400$\\
20 & Intermittent contact and collective problem solving \citeapp{bernstein2018intermittent} & 8 agents $\times$ 10 rounds & intermittent $-$ continuous $<0$ & \same{5.0}/\zero{0.0}/\opp{0.0} & 1.000 & 0.000 & 1.000\\
21 & Group recombination and cumulative design \citeapp{derex2016partial} & 8 agents $\times$ 10 rounds & partial $-$ complete $>0$ & \same{2.4}/\zero{0.0}/\opp{2.6} & 0.475 & 0.525 & $-0.050$\\
22 & Cultural-market popularity accumulation \citeapp{salganik2006experimental} & 8 agents $\times$ 8 rounds & visible $-$ hidden counts $>0$ & \same{2.5}/\zero{0.0}/\opp{2.5} & 0.500 & 0.500 & 0.000\\
23 & Multi-round signals and action cascades \citeapp{anderson1997information} & 8 agents $\times$ 8 rounds & visible $-$ hidden history $>0$ & \same{2.4}/\zero{1.0}/\opp{1.6} & 0.475 & 0.325 & 0.150\\
24 & Market-price feedback and convergence \citeapp{smith1962experimental} & 8 agents $\times$ 8 rounds & public $-$ private tape $>0$ & \same{3.8}/\zero{0.0}/\opp{1.2} & 0.750 & 0.250 & 0.500\\
25 & Stable identity in labor markets \citeapp{brown2004relational} & 8 agents $\times$ 8 rounds & stable $-$ fresh identity $>0$ & \same{4.5}/\zero{0.0}/\opp{0.5} & 0.900 & 0.100 & 0.800\\
26 & Fulfillment reputation in online trade \citeapp{bolton2004effective} & 8 agents $\times$ 8 rounds & public $-$ private experience $>0$ & \same{4.5}/\zero{0.0}/\opp{0.5} & 0.900 & 0.100 & 0.800\\
27 & Weak-link coordination and group size \citeapp{vanhuyck1990tacit} & 8 agents $\times$ 8 rounds & pairs $-$ whole group $>0$ & \same{5.0}/\zero{0.0}/\opp{0.0} & 1.000 & 0.000 & 1.000\\
28 & Cumulative mitigation and collective risk \citeapp{milinski2008collective} & 6 agents $\times$ 10 rounds & high $-$ low risk $>0$ & \same{3.1}/\zero{0.2}/\opp{1.6} & 0.625 & 0.325 & 0.300\\
29 & Communication organization and team aggregation \citeapp{leavitt1951effects} & 8 agents $\times$ 8 rounds & star $-$ ring $>0$ & \same{0.4}/\zero{4.1}/\opp{0.5} & 0.075 & 0.100 & $-0.025$\\
30 & Intergroup competition and coordination \citeapp{bornstein2002effect} & 8 agents $\times$ 8 rounds & competition $-$ independent $>0$ & \same{2.5}/\zero{1.2}/\opp{1.2} & 0.500 & 0.250 & 0.250\\
\bottomrule
\end{longtable}
}

\twocolumn
\bibliographystyleapp{ACM-Reference-Format}
\bibliographyapp{ref-sup}

\end{document}